\documentclass[journal,compsoc]{IEEEtran}

\usepackage{graphicx}
\usepackage{ragged2e}
\usepackage{cite}
\usepackage[colorlinks, citecolor=blue]{hyperref}
\usepackage{pifont}
\usepackage{multirow}
\usepackage{listings}
\usepackage{xcolor}
\usepackage{float}
\usepackage{xspace}
\usepackage{hyperref}
\usepackage{wasysym}

\newfloat{lstfloat}{htbp}{lo1}
\floatname{lstfloat}{Listing}

\newcommand{\sysname}{\textsc{MaTEE}\xspace}
\newcommand{\ssecref}[1]{\mbox{\S\textcolor{green}{\ref{#1}}}\xspace}
\newcommand{\RNum}[1]{\uppercase\expandafter{\romannumeral #1\relax}}

\definecolor{clr-background}{RGB}{255,255,255}
\definecolor{clr-text}{RGB}{0,0,0}
\definecolor{clr-string}{RGB}{163,21,21}
\definecolor{clr-namespace}{RGB}{0,0,0}
\definecolor{clr-preprocessor}{RGB}{128,128,128}
\definecolor{clr-keyword}{RGB}{0,0,255}
\definecolor{clr-type}{RGB}{43,145,175}
\definecolor{clr-variable}{RGB}{0,0,0}
\definecolor{clr-constant}{RGB}{171,0,118}
\definecolor{clr-comment}{RGB}{0,128,0}

\lstdefinestyle{c-style}{
  language=C,
  sensitive=true,
  basicstyle=\linespread{0.8}\footnotesize\ttfamily,
  rulesepcolor=\color{gray},
  numberstyle=\scriptsize\color{gray},
  keywordstyle=\bfseries,
  keywordstyle=[9]\color{blue},
  stringstyle=\color{clr-constant},
  identifierstyle=\color{clr-variable},
  commentstyle=\color{clr-comment},
  emphstyle=\color{clr-type},
  escapeinside={/*!}{!*/},
}

\begin{document}
\title{\sysname: Efficiently Bridging the Semantic Gap in TrustZone via Arm Pointer Authentication}

\author{Shiqi~Liu,
        Xiang~Li,
        Jie~Wang,
        Yongpeng~Gao,
        Jiajin~Hu
    
        \IEEEcompsocitemizethanks{
            \IEEEcompsocthanksitem Shiqi~Liu,
        Jie~Wang,
        Yongpeng~Gao, and
        Jiajin~Hu are with Hubei Key Laboratory of Distributed System Security, Hubei Engineering Research Center on Big Data Security, School of Cyber Science and Engineering, Huazhong University of Science and Technology, Wuhan, 430074, China, also with JinYinHu Laboratory, Wuhan, 430040, China. \\E-mail: \{shiqiliu,wangjie\_s,sternen\_hust,hujj\_cse01\}@hust.edu.cn.
            \IEEEcompsocthanksitem     Xiang~Li is with Research Center for Basic Theories of Intelligent Computing, Research Institute of Basic Theories, Zhejiang Laboratory, Hangzhou, 311100, China. E-mail: sean.lixiang97@gmail.com.
          \IEEEcompsocthanksitem Corresponding author: Jie~Wang. 
            \IEEEcompsocthanksitem This work is partially supported by the National Natural Science Foundation of China under Grants 62202194, and sponsored by CCF-Huawei Populus Grove Fund.
          }
          \thanks{}
}

\markboth{IEEE Transactions on Dependable Secure Computing}
{Shell \MakeLowercase{\textit{et al.}}: Bare Demo of IEEEtran.cls for Computer Society Journals}

\IEEEtitleabstractindextext{
\begin{abstract}
   \justifying
    Trusted Execution Environments (TEEs) employ hardware-based isolation mechanisms to safeguard the confidentiality and integrity of sensitive code and data. One such prevalent implementation is Arm TrustZone, which partitions the system into the secure and normal (non-secure) worlds. However, this partitioning results in the secure world having very limited visibility into the operating information of the normal world, creating a semantic gap between these two worlds. Specifically, the secure world lacks an effective user identity authentication when receiving data requests from the normal world. Consequently, malicious Client Applications (CAs) in the normal world can deceive Trusted Applications (TAs) in the secure world by utilizing elaborate request parameters, compromising the sensitive data stored by other CAs. We systematically classify these Semantic Gap Vulnerabilities (SGVs) and propose a mate system for the TEE called \sysname to defend against SGVs. \sysname utilizes Arm Pointer Authentication (PA) to bind each request to the corresponding CA's identity and then verifies the identity when the CA accesses sensitive data, thereby preventing malicious request forgery. In particular, \sysname isolates sensitive data of different CAs without modifying existing CAs and TAs. Our evaluation demonstrates that \sysname successfully defends against SGVs with a minimal runtime overhead (2.19\%).
\end{abstract}

}

\maketitle

\IEEEdisplaynontitleabstractindextext

\IEEEpeerreviewmaketitle

\IEEEraisesectionheading{\section{Introduction}\label{sec:introduction}}

\IEEEPARstart{T}{raditional} security mechanisms (e.g., software sandboxing~\cite{sehr2010adapting},~\cite{morrisett2012rocksalt},~\cite{narayan2020retrofitting},~\cite{johnson2021trust},~\cite{bosamiya2022provably}) cannot defend against attacks from privileged software (e.g., OS or hypervisor). To counter this, Arm company introduced TrustZone~\cite{arm2009security} starting from the Armv6 architecture. 
The Rich OS (e.g., Linux or Android) and CAs operate in the normal world, while TEE OS (e.g., OP-TEE~\cite{mcgillion2015open} or QSEE~\cite{khalid2022vulnerability}) manages TAs in the secure world. TAs are developed by TEE vendors to handle sensitive operations and data. To access the secure functionalities of TAs, CAs need to transmit requests and supply parameters to the secure world. 

The isolation design of TrustZone is unidirectional. When the CPU operates in a non-secure state, it inhibits CAs and the Rich OS from accessing memory, cache, or peripheral hardware devices allocated for the secure world. Conversely, TEE OS retains unrestricted access permissions to resources in both the secure and normal worlds. However, this asymmetry in access control allows attackers to exploit vulnerabilities~\cite{shen2015exploiting},~\cite{Beniamini2016CVE},~\cite{rosenberg2014reflections},~\cite{berard2018kinibi},~\cite{Komaromy2018Unbox} in the secure world to launch confused deputy attacks~\cite{hardy1988confused} against the normal world. 
Recently, some novel attacks have been discovered during cross-world interactions, which are identified as Semantic Gap Vulnerabilities (SGVs)~\cite{machiry2017boomerang},~\cite{suciu2020horizontal}. Although the secure world can access the normal world, it often has limited visibility due to the semantic gap, making it difficult to verify the identities of CAs and request parameters. This gap is exploited when malicious CAs send forged requests to TAs, leading to unauthorized access to sensitive data within TEE; address spaces of other CAs, or potentially even the Rich OS.

We classify SGVs into Type-\RNum{1} SGVs (based on pointers) and Type-\RNum{2} SGVs (based on resource IDs). \textbf{Type-\RNum{1} SGVs}: TrustZone allows CAs to share pointers or data with the secure world. Malicious CAs exploit Type-\RNum{1} SGVs by transmitting pointers pointing to other CAs or the Rich OS, confusing the TA to assist in achieving arbitrary address access to the normal world. 
Although the Rich OS performs pointer sanitization (PTRSAN) to prevent out-of-bounds access, there are two methods to bypass PTRSAN: the first method exploits a flaw in implementing PTRSAN by checking only the base address of the access region but lacking inspection for the offset, and the second method adopts disguising pointers as non-pointer data to bypass inspection (i.e., type confusion).
\textbf{Type-\RNum{2} SGVs}: 
When processing requests from a CA, a TA stores sensitive CA-related data in the TEE region. Subsequently, it allocates a resource ID, such as a shared memory ID, to facilitate future data access. TrustZone lacks proper authentication for different CAs connected to the same TA. Any CA with the correct ID can access the sensitive data of other CAs without limitations. Moreover, sensitive data can also reside in the global variables of TAs. These variables are available for multiple TA sessions and can be accessed through CA requests.

Cooperative Semantic Reconstruction (CSR)~\cite{machiry2017boomerang} is proposed to mitigate type confusion in Type-\RNum{1} SGVs. When parameters are passed from the Rich Execution Environment (REE) to the TEE, the REE requests the type of parameters from the TEE for PTRSAN. However, CSR cannot address Type-\RNum{2} SGVs, and its reliance on multiple cross-world interactions increases communication overhead. In addition to CSR, Type-\RNum{1} SGVs have comprehensive solutions in existing TEEs. For instance, in OP-TEE~\cite{mcgillion2015open}, each parameter transmission requires an additional parameter type, which undergoes rigorous verification by the TA. The PTRSAN in the TEE driver checks the validity of the base address and offset of each pointer parameter. Moreover, all pointers must point to predefined shared memory, preventing arbitrary address access and rendering pointer-based attacks ineffective. Therefore, for Type-\RNum{1} SGVs, we reuse the existing solutions present in OP-TEE.

\textbf{Key Ideas}. In this paper, we propose \sysname, an approach against Type-\RNum{2} SGVs. To the best of our knowledge, our work represents the first attempt at addressing Type-\RNum{2} SGVs. In addition, we find new triggering strategies and inter-thread data theft for Type-\RNum{2} SGVs (\ssecref{subsec:problem}). The resource IDs used to access TEE resources are not bound to the identity of the CA. Worse, these IDs are easily guessed and obtainable. There are two common approaches to address this issue: access control~\cite{samarati2000access} and capabilities~\cite{carter1994hardware},~\cite{watson2015cheri},~\cite{filardo2020cornucopia},~\cite{yu2023capstone}. The access-control-based approach requires recording the explicit access policies for the CA identity and corresponding TEE resources. In contrast, the capability-based design does not necessitate explicit access control policies. The capabilities offer an unforgeable token used for access control, which is implicitly contained within the data transmission between the accessor and the accessed. \sysname leverages Arm Pointer Authentication (PA) to provide the capability, signing resource IDs with the CA identity during allocation. The signature is verified when CAs utilize these resource IDs to access TEE resources.

We place the Pointer Authentication Code (PAC), the signature signed by PA, in the higher-order bits of the 64-bit resource ID (originally 32 bits and extended in our system). This PAC can be automatically propagated by copying or moving the resource ID, eliminating the need for extra memory allocation and tracking. The TEE driver provides the identity of the CA and transmits it to the secure world for resource ID signing and verification. Therefore, the secure world no longer requires managing any identity of the CA. \sysname achieves process- and thread-level isolation for CA resources and supports legal sharing for CA threads. The capability-based design can enforce fine-grained isolation within the CA by configuring different threads with different privileges. This ensures that sensitive operations are restricted to authorized threads only, thereby minimizing the potential impact of a compromised thread.

To defend against Type-\RNum{2} SGVs, we face the following challenges. \textbf{Challenge 1: How does the secure world differentiate CA processes and threads?} 
Due to the semantic gap, the secure world cannot identify which CA process or thread initiates a request. To address this, we modify the TEE driver to append the identity of the CA process to each request (\ssecref{subsec:identity_management}). The CA process ID serves as the PA context used for PAC generation (\ssecref{subsec:pac_sign}), ensuring resource isolation among different CA processes and resource sharing within threads of the same CA process. Since each thread interacts with the secure world via a unique TA session, we utilize the session ID from the TEE OS to differentiate CA threads. Resource isolation is achieved between threads by combining the CA process ID and the session ID as the PA context (\ssecref{subsec:pac_sign}).
\textbf{Challenge 2: How to defend against brute-force cracking and replay attacks on signatures?} 
The bits allocated to the PAC are inherently limited, making it susceptible to brute-force attacks. To address this, we introduce an exception-handling module (\ssecref{subsec:exception}). If signature verification fails, this module raises an exception and terminates the malicious CA. 
In addition to brute-force attacks, if the CA still has sensitive data stored in the TEE upon exiting, a malicious CA might be assigned the same CA ID, potentially initiating a replay attack. Instead of using the process ID directly as the PA context, we enhance the system to generate a 32-bit random number that serves as the CA process identity (\ssecref{subsec:identity_management}). Furthermore, a timestamp is added to the session ID to prevent replay attacks initiated by malicious threads.
\textbf{Challenge 3: How to ensure the consistency of PAC keys used for signing and verifying in the secure world?} Each trusted thread in the secure world has a different PAC key. However, requests from the same CA may be executed by different trusted threads, rendering the keys used for signing and verifying for the same CA unsynchronized. To address this, we modify the trusted thread scheduling module (\ssecref{subsec:thread_management}) of the TEE OS to ensure that a single trusted thread exclusively handles requests from the same CA.

We develop two prototypes of \sysname, deploying the functional prototype on the Arm Fixed Virtual Platform (FVP)~\cite{ARMFVP} and the performance prototype on the Rock Pi 4B development board (\ssecref{sec:prototype}). We conduct the security evaluation against Horizontal Privilege Escalation (HPE)~\cite{suciu2020horizontal}, \textsc{Boomerang}~\cite{machiry2017boomerang}, and other SGVs. The results show that \sysname is immune to these vulnerabilities (\ssecref{subsec:mitigation}). To evaluate the performance of \sysname, we adopt the \texttt{xtest}~\cite{OPTEExtest} microbenchmarks (\ssecref{subsec:microbenchmarks}) and port three real-world applications (\ssecref{subsec:applications}), including \texttt{AVB}~\cite{Wiklander2018avb} for verifying Android boot, \texttt{Trusted keys}~\cite{Garg2020trustedkeys} for sealing keys, and \texttt{DarkneTZ}~\cite{mo2020darknetz}, a deep neural network for classification tasks. Our experimental results show that \sysname has an average runtime overhead of 2.19$\%$.

We make the following contributions:

\begin{itemize}
  \item We propose \sysname, a defense mechanism that leverages Arm PA to counter SGVs. \sysname supports the isolation and legal sharing of TEE resources between CAs.
  \item We categorize SGVs and demonstrate that our defense is effective against them. We build and release a test suite for evaluating these vulnerabilities.
  \item We develop and open-source a prototype of \sysname\footnotemark. This system is compatible with existing CAs and TAs and introduces only 984 LoCs to the TCB.
  \item We test \sysname on both simulators and real-world platforms. The results demonstrate that \sysname incurs minimal overhead.
\end{itemize}

\footnotetext{\url{https://github.com/erhade/MaTEE}}

\section{Background} \label{sec:background}
In this section, we provide an overview of TrustZone (\ssecref{subsec:TrustZone}), Semantic Gap Vulnerabilities (\ssecref{subsec:semantic-gap}), and Arm Pointer Authentication (\ssecref{subsec:armPA}).
\subsection{TrustZone for Armv8 Architecture} \label{subsec:TrustZone}
Arm TrustZone is a security extension designed to provide isolated protection for sensitive code and data. We focus on Armv8-A, the first Arm architecture to offer AArch64 support. In this architecture, each physical core splits into two virtual cores. The state of the current Arm core is indicated by the Non-secure (NS) bit on the AXI bus. The system employs the \texttt{SMC} (Secure Monitor Call) instruction to switch between the secure and non-secure states. TrustZone ensures that resources in the secure world are shielded from the normal world. Memory isolation is managed through the TrustZone Address Space Controller (TZASC), and peripheral devices are safeguarded by configuring the TrustZone Peripheral Controller (TZPC).

\begin{figure}[!t]
  \centering
  \includegraphics[width=2.3 in]{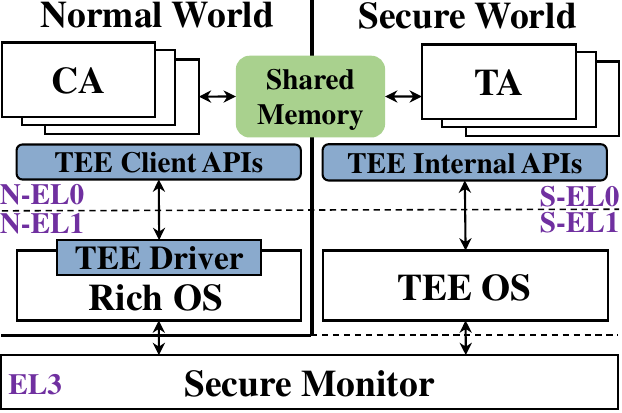}
  \caption{Armv8-A TrustZone software architecture.}
  \label{fig:TrustZoneSoftware}
\end{figure}

The software architecture of Armv8-A TrustZone is illustrated in Fig.~\ref{fig:TrustZoneSoftware}. Armv8-A contains four exception levels (EL0-EL3) and two security states (secure and non-secure states)~\cite{ARM2022exception}. The REE comprises CAs at EL0 and the Rich OS at EL1. Correspondingly, the TEE comprises TAs at EL0 and the TEE OS at EL1. EL2 is reserved for a hypervisor (this paper does not deal with hypervisors), while EL3 is allocated for the Trusted Firmware used for security state switching (i.e., Secure Monitor). GlobalPlatform standardizes the TrustZone interfaces for the REE and TEE. The encapsulated interface handling the interaction between the CA and the TEE driver is called the TEE Client APIs~\cite{GlobalPlatform2010ClientAPI}, while the interface managing the interaction between the TA and TEE OS is termed TEE Internal APIs~\cite{GlobalPlatform2021API}. Shared memory enables data transfer between the CA and TA. This shared memory can be established by registering existing memory or newly allocated memory from CA.

Some endpoints (e.g., embedded devices) lack the TrustZone architecture and therefore must rely on server-based TEE services. On these servers, both the TA and the CA operate. Upon receiving a request from an endpoint, the CA initiates a new thread to manage it. This thread, when connecting to a TA, triggers the establishment of a TA session, ensuring a direct correspondence between CA threads and TA sessions. TAs function in various modes to balance security and performance requirements. In single-instance mode, a single TA instance handles multiple sessions simultaneously, saving resources but offering less data isolation between sessions. In contrast, multi-instance mode assigns a unique TA instance to each session, enhancing data isolation at the cost of higher memory usage.

\begin{figure}[!t]
  \centering
  \includegraphics[width=3.5 in]{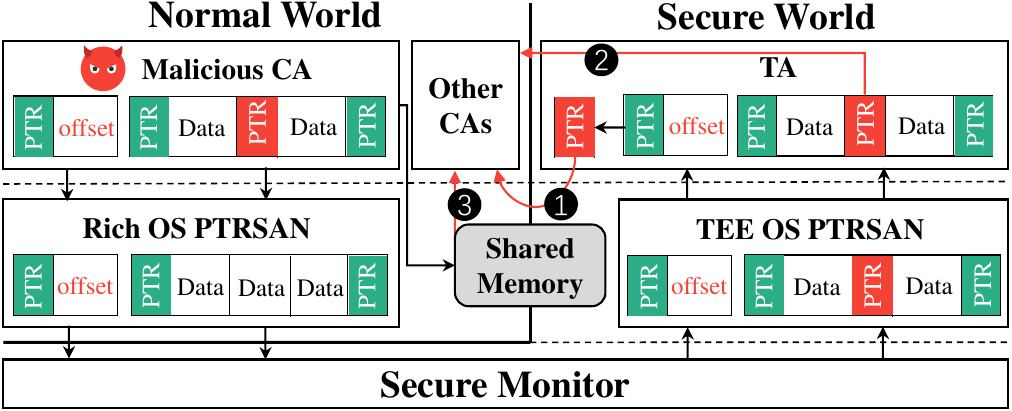}
  \caption{\textsc{Boomerang} attacks. \ding{182} PTRSAN Bypass: no offset check, leading to pointer out-of-bounds. \ding{183} Type Confusion: pointer mistaken as data, bypassing PTRSAN. \ding{184} Shared Memory Hijacking: Unauthorized access via forged shared memory ID.}
  \label{fig:Boomerang}
\end{figure}

\begin{figure}[!t]
  \centering
  \includegraphics[width=3.5 in]{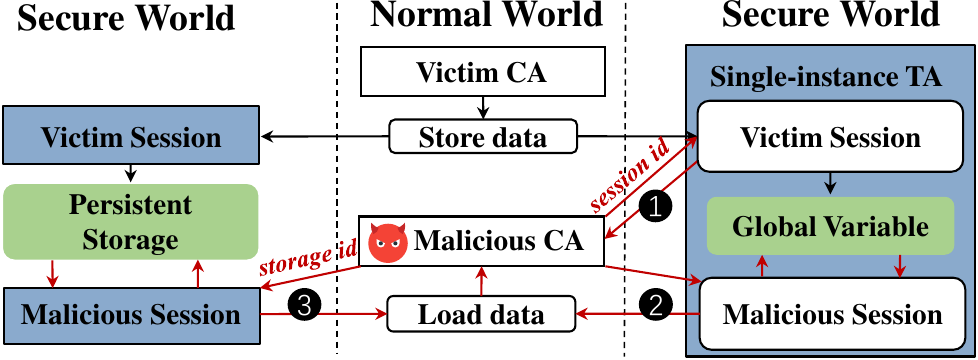}
  \caption{HPE Attacks Leading to Data Leakage. \ding{182} Directly access the victim session using a forged session ID. \ding{183} Data retrieval from global variables across sessions within a single-instance TA. \ding{184} Unauthorized access to data in persistent storage via a forged storage ID.}
  \label{fig:HPE}
\end{figure}

\subsection{Semantic Gap Vulnerabilities} \label{subsec:semantic-gap}

Semantic Gap Vulnerabilities (SGVs) arise from insufficient semantic information, 
making it challenging for the secure world to validate request parameters and requester identity from the normal world. SGVs covered in previous research include \textsc{Boomerang}~\cite{machiry2017boomerang} and Horizontal Privilege Escalation (HPE)~\cite{suciu2020horizontal}.

BOOMERANG attacks can be divided into three primary types, as illustrated in Fig.~\ref{fig:Boomerang}:

\ding{182} \textbf{PTRSAN Bypass}: This type of vulnerability capitalizes on the flawed implementation of PTRSAN. When a CA utilizes a base address combined with an offset as a command parameter, PTRSAN in the Rich OS only verifies the base address within the CA's address space. Upon successful verification, the base address is translated to a physical address. When the TA accesses this address with the appended offset, it may inadvertently relay sensitive data to a malicious CA.

\ding{183} \textbf{Type Confusion}: This attack allows the malicious CA to disguise a pointer as non-pointer data when sending it to the TEE. On the REE side, as message structures are TA-defined, the Rich OS faces difficulty identifying parameters that are pointers needing sanitization. On the TEE side, the TEE OS's PTRSAN can only check if the pointer incorrectly points to the secure world's memory. Malicious CAs can exploit this semantic gap to access the address spaces of other CAs or the Rich OS.

\ding{184} \textbf{Shared Memory Hijacking}: This attack involves a malicious CA manipulating the shared memory ID of a victim CA. Due to the lack of strict isolation, the malicious CA can access the shared memory of other CAs, potentially exposing or corrupting the victim CA's sensitive data.

In HPE attacks, malicious CAs can exploit stateful TAs, which preserve CA data across session states (e.g., global variables) or persistent storage, to leak or corrupt sensitive data. Fig.~\ref{fig:HPE} illustrates three ways to exploit HPE vulnerabilities:

\ding{182} \textbf{TA Session Hijacking}: Malicious CAs can forge requests using the session ID of a victim CA, allowing unauthorized access to the TA session. This enables them to steal sensitive data or execute unauthorized operations.

\ding{183} \textbf{Global Variable Hijacking:} In single-instance mode, multiple CAs connect to the same TA instance with multiple sessions, and global variables in the TA instance can be accessed without restriction by multiple CAs through sessions. Malicious CAs can exploit this to leak session data after a victim CA stores sensitive data in the global variables.

\ding{184} \textbf{Persistent Storage Hijacking:} 
Coarse-grained access control allows multiple sessions of the same TA (including both single-instance and multi-instance modes) to share storage resources. A malicious CA can bypass access control by using the storage ID of a victim CA as a request parameter, leading to data leakage in persistent storage.

We also find new attack variants of SGVs, specifically those that compromise TA heap addresses and opaque handles. Opaque handles are abstract references to TEE resources, keeping their internal structure hidden from the caller. We provide a detailed categorization of SGVs in \ssecref{subsec:problem}.

\begin{figure}[!t]
  \centering
  \includegraphics[width=3 in]{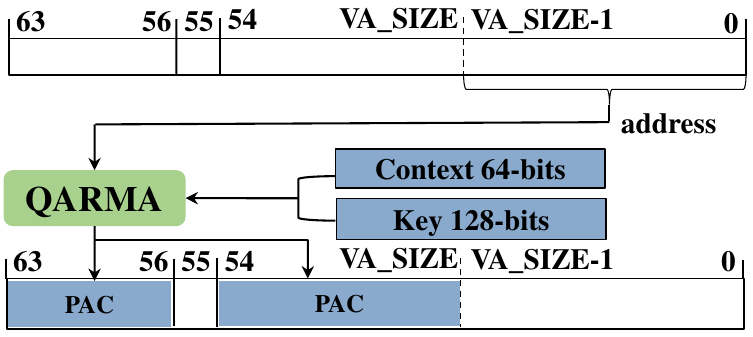}
  \caption{Signing pointers with PAC.}
  \label{fig:PAC}
\end{figure}

\subsection{Arm Pointer Authentication} \label{subsec:armPA}
Arm introduces Pointer Authentication (PA)~\cite{Qualcomm2017PA} from Armv8.3-A, using it to maintain pointer integrity. Specifically, PA signs critical data (like return addresses or function pointers) and verifies the signature prior to usage for secure control flow integrity (CFI). In Armv8.3-A, PA only sets the error bit in case of pointer authentication failure and still signs the pointer with the error bit, which leaves it susceptible to brute-force and signature reuse attacks. To counter signature reuse attacks, Armv8.6-A employs an XOR operation between signatures and high pointer bits~\cite{farkhani2021ptauth}. In addition, to prevent brute-force attacks, the PA on Armv8.6-A directly throws an exception when pointer authentication fails. \sysname is constructed on the FVP simulator that supports the Armv8.3-A PA feature. We also incorporate additional software layers in our design to safeguard against brute-force and signature reuse attacks.

\begin{table*}[!th]
  \caption{SGVs in OP-TEE, QSEE, and Kinibi. Symbol \CIRCLE~~indicates that the platform can resist SGVs between CA threads and processes, symbol \LEFTcircle~~indicates that the platform can only resist SGVs between CA processes, and symbol \Circle~~indicates that the platform does not implement any defenses against SGVs. In TA, $M$ stands for multi-instance mode, and $S$ stands for single-instance mode.}
  \label{tab:attacks}
  \centering

  \begin{tabular}{l|l|c|c|c|c|c|l}
  \hline

  \hline
  \multicolumn{2}{c|}{\textbf{Semantic Gap Vulnerabilities (SGVs)}} & \textbf{OP-TEE} & \textbf{QSEE} & \textbf{Kinibi} & \textbf{TA} & \textbf{Type} & \textbf{Previous research}\\
  \hline

  \hline
    \multirow{2}{*}{Type-\RNum{1} SGVs} &  Bypass pointer sanitization & \CIRCLE & \Circle & \Circle & $M\&S$ & CWE-200 & \textsc{Boomerang}~\cite{machiry2017boomerang}\\

    & Confuse pointer type & \CIRCLE & \CIRCLE & \CIRCLE &  $M\&S$  & CWE-200 & \textsc{Boomerang}~\cite{machiry2017boomerang}\\
  \hline
    \multirow{6}{*}{Type-\RNum{2} SGVs} & Hijack shared memory & \LEFTcircle & \LEFTcircle & \LEFTcircle & $M\&S$ & CWE-639 & \textsc{Boomerang}~\cite{machiry2017boomerang}\\

     & Hijack TA sessions & \LEFTcircle & \Circle & \Circle & $M\&S$ & CWE-639 & HPE~\cite{suciu2020horizontal} \\

     & Compromise global variables & \Circle & \Circle & \Circle & $S$ & CWE-862 & HPE~\cite{suciu2020horizontal}\\

     & Compromise TA heap & \Circle & \Circle & \Circle & $S$ & CWE-862 &\\

     & Compromise by opaque handles & \Circle & \Circle & \Circle & $M\&S$ & CWE-732 &\\

     & Hijack persistent storage & \Circle & \Circle & \Circle & $M\&S$ & CWE-732 & HPE~\cite{suciu2020horizontal}\\
  \hline

  \hline
  \end{tabular}
\end{table*}

Arm PA utilizes the QARMA~\cite{avanzi2017qarma} algorithm to generate a 64-bit signature integrating a pointer value, a 64-bit context (known as a modifier), and a 128-bit key. This signature is then truncated to create a PAC, as illustrated in Fig.~\ref{fig:PAC}. However, a conflict arises with PAC due to using $Xn\left[ 63:56 \right]$ by the Memory Tag Extension (MTE). To resolve this issue, Arm specifies that PAC uses $Xn\left[ 54:\mathrm{VA}\_\mathrm{SIZE} \right]$ when MTE is enabled, while $Xn\left[ 54:\mathrm{VA}\_\mathrm{SIZE} \right]$ and $Xn\left[ 63:56 \right]$ are utilized when MTE is disabled. The Linux kernel uses a 48-bit virtual address space, limiting the PAC to 15 bits. TEE OS uses a 32-bit virtual address, allowing the PAC to extend up to 31 bits when MTE is disabled.

Arm PA provides five root keys, which include two keys each for instructions and data (specifically, Instruction Key A, Instruction Key B, Data Key A, and Data Key B), along with a generic key. All PAC key registers are situated at EL1 and remain inaccessible from user space. The Rich OS and TEE OS manage two sets of keys for the user mode and the kernel mode, loading the appropriate keys into the PAC key registers to activate them in different modes. Despite this, user-space applications can still execute PA instructions to generate a PAC using the user mode PAC key. PAC-related instructions include \texttt{pac*} for PAC generation, \texttt{aut*} for PAC authentication, and \texttt{xpac*} to remove PAC from the high bits of the pointer. In contrast to the user mode, which can use all five root keys, the kernel mode only utilizes Instruction Key A. In this paper, we employ the \texttt{pacia} instruction (using Instruction Key A) to sign the resource ID and the \texttt{autia} instruction (also using Instruction Key A) to verify the signature. Additionally, PAC allows for context specification, facilitating unique signatures across different computing contexts. We use these contexts to distinguish between various CA processes and threads (\ssecref{subsec:pac_sign}).

\section{Overview} \label{subsec:overview}
In this section, we first introduce \sysname's threat model (\ssecref{subsec:model}) and then summarize six Type-\RNum{2} SGVs (\ssecref{subsec:problem}). Following that, we outline the system's implementation requirements and the design of its main components (\ssecref{subsec:requirements}) that can address these attacks.

\subsection{Threat Model} \label{subsec:model}

\sysname trusts the secure world software stack. Given that SGVs involve attacks by normal world CAs against other CAs and the Rich OS, mitigation of these attacks depends on the Rich OS to supply essential semantic information; thus, \sysname extends its trust to the Rich OS as well. Additionally, \sysname trusts the hardware, assuming Arm TrustZone conforms to its specifications.

We focus on defending against attacks that exploit Type-\RNum{2} SGVs, including \textsc{Boomerang}~\cite{machiry2017boomerang}, HPE~\cite{suciu2020horizontal}, and other potential attacks related to TA-returned heap addresses and opaque handles. We assume that attackers can compromise a CA process or thread. This attacker would have permission to communicate with a target TA and acquire resource IDs, such as shared memory IDs and storage IDs, from other CA processes or threads within the same CA, potentially triggering Type-\RNum{2} SGVs. We leverage Arm TrustZone technology for robust isolation and Arm PA to ensure integrity protection without necessitating rewrites of CA or TA code.

Our scope does not include side-channel attacks, such as PACMAN~\cite{ravichandran2022pacman}, or denial-of-service attacks. Additionally, we do not address physical attacks targeting internal hardware components, such as cold-boot attacks~\cite{yitbarek2017cold} or bus snooping attacks~\cite{lee2020off}.

\subsection{Problem Statement} \label{subsec:problem}

We conduct an in-depth study of the principles behind SGVs and perform validation tests. In the process, we classify and discover new attack variants (i.e., exploits by TA-returned heap addresses and opaque handles) using known Common Weakness Enumeration (CWEs)~\cite{MITRE2023cwe}. We systematically categorize all SGVs found into Type-\RNum{1}, based on pointer manipulation, and Type-\RNum{2}, based on resource ID attacks (Table~\ref{tab:attacks}). Type-\RNum{1} SGVs encompass pointer sanitization bypass and pointer type confusion, leading to \textsc{Boomerang} attacks (\ssecref{subsec:semantic-gap}). These attacks fall under CWE-200 (Exposure of Sensitive Information to an Unauthorized Actor), resulting in sensitive data leakage from other CAs or the Rich OS to unauthorized CAs (Fig.~\ref{fig:Boomerang}). While QSEE and Kinibi's pointer sanitization can be bypassed, OP-TEE effectively defends against Type-\RNum{1} SGVs through strict type checks and by enforcing pointers to reference shared memory, eliminating the need for additional defenses against Type-\RNum{1} SGVs in \sysname based on the OP-TEE platform.

Type-\RNum{2} SGVs encompass a variety of vulnerabilities: shared memory hijacking (see \textsc{Boomerang} attacks in \ssecref{subsec:semantic-gap}); TA session hijacking (see HPE attacks in \ssecref{subsec:semantic-gap}); compromising global variables and persistent storage (see HPE attacks in \ssecref{subsec:semantic-gap}); and vulnerabilities related to TA-returned heap addresses and opaque handles (as detected in our tests). Notably, global variables and heap vulnerabilities predominantly target single-instance TAs, while the rest can affect both single and multi-instance TAs. We summarize the SGVs for three platforms: OP-TEE, QSEE, and Kinibi (Table~\ref{tab:attacks}). Both QSEE and Kinibi offer only partial defenses against shared memory hijacking. Meanwhile, OP-TEE drivers combat TA session and shared memory hijacking by retaining unique session ID lists for each CA process and associating shared memory with its context. However, OP-TEE cannot counter session or shared memory hijacking by malicious threads within the same CA. Overall, these platforms lack comprehensive defense mechanisms against Type-\RNum{2} SGVs. We classify Type-\RNum{2} SGVs into three CWE categories, each containing two specific attacks. Here, we introduce these attacks and our defense measures against them:

\textbf{CWE-639: Authorization Bypass Through User-Controlled Key.} This attack arises from guessable IDs that the TEE returns, specifically for shared memory and TA sessions. As authentication determines the validity of data only based on the ID, malicious CAs forge these IDs to bypass authentication, access sensitive data in shared memory, or hijack other CAs' sessions for unauthorized operations. To counter authorization bypass attacks (\ssecref{subsec:auth_bypass}), we enhance the TEE driver in the Rich OS to use PAC for ID signing during allocation, incorporating the CA's identity. The signature is verified when accessing resources using IDs.

\textbf{CWE-862: Missing Authorization.} A single-instance TA serves multiple CAs, and sensitive data may be stored in TA global variables and the heap (with the TA passing the heap address to CAs as a token for accessing sensitive data). The lack of isolation for these variables allows malicious CAs to cause data leakage and corruption. To counter missing authentication attacks (\ssecref{subsec:missing_auth}), we disable global variables through static analysis and propose two alternative methods that use PAC to share data globally. To prevent unauthorized access when the TA returns the heap address to CAs, we introduce a new data type specifically designed for data transfer between the two worlds, utilizing PAC to ensure integrity protection and identity authentication.

\textbf{CWE-732: Incorrect Permission Assignment for Critical Resource.} Due to insufficient fine-grained isolation, malicious CAs can access the persistent storage of other CAs within the TEE OS. Additionally, when TAs request sensitive resources (e.g., keys), the TEE OS issues opaque handles, such as the \texttt{OperationHandle} listed in Table~\ref{tab:OpaqueHandles}, which are then returned to the CAs. Malicious CAs can acquire these handles from other CAs and potentially access sensitive data. To counter incorrect permission assignment attacks (\ssecref{subsec:permission}), we have revised all TEE Internal APIs that involve opaque handles. We now incorporate PAC signing with CA identity during opaque handle allocation and validate these signatures during usage. However, merely signing opaque handles is insufficient to thwart storage hijacking attacks, as attackers can request new handles using an existing storage ID to circumvent the verification process. To address this, we have overhauled the storage management mechanism within the TEE OS to ensure that accessing persistent storage with the same storage ID consistently returns the existing opaque handle, rather than creating a new one.

\subsection{Requirements and Framework} \label{subsec:requirements}
To achieve an effective and user-friendly solution for Type-\RNum{2} SGVs, \sysname aims to meet the following goals:

\begin{enumerate}
    \item \textbf{Scalability}: Defend against existing SGVs and have good scalability for potential future SGVs.
    \item \textbf{Compatibility}: Ensure compatibility with existing CAs and TAs, avoiding application rewrites.
    \item \textbf{Security}: Minimize modifications to the TCB and avoid architectural changes to the Rich OS or the TEE OS.
    \item \textbf{Efficiency}: Minimize performance overhead to ensure optimal REE side call performance.
\end{enumerate}

\begin{figure}[!t]
  \centering
  \includegraphics[width=3.3 in]{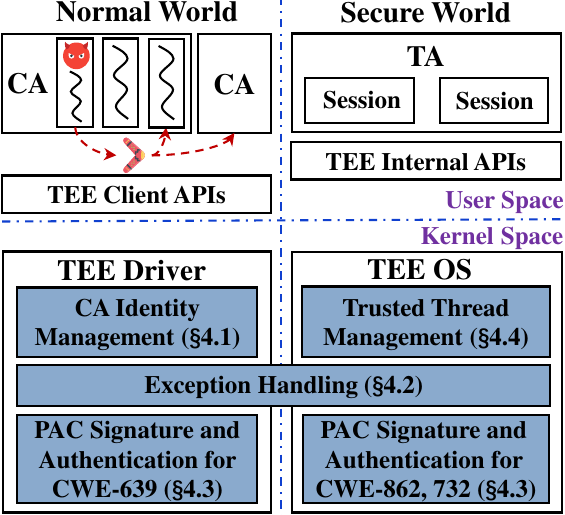}
  \caption{The Architecture of \sysname (The dark areas are the modules of our system).}
  \label{fig:MaTEE}
\end{figure}

Fig.~\ref{fig:MaTEE} shows the architecture of \sysname. During TEE resource allocation for CWE-639, the PAC module (\ssecref{subsec:pac_sign}) in the TEE driver signs the resource ID allocated on the REE side, including shared memory and session IDs. For CWE-862 and CWE-732, the PAC module within the TEE OS signs the resource ID allocated by the TEE, and this signed ID is then forwarded back to the CA through the TEE driver. When a CA initiates a request for resource allocation to the TEE using the \texttt{SMC} instruction, the CA Identity Management (\ssecref{subsec:identity_management}) in the TEE driver (part of the Rich OS) appends each CA's identity (serving as the PA context) as parameters to the \texttt{SMC} instruction. This identity is then used by the PAC module within the TEE OS. For later access to TEE resources, the TEE driver's PAC module verifies the signature for CWE-639, while the TEE OS's PAC module handles verification for CWE-862 and CWE-732. The cross-world exception handling module (\ssecref{subsec:exception}) terminates the malicious CA if authentication fails. Additionally, the trusted thread management (\ssecref{subsec:thread_management}) in the TEE OS ensures that each CA's request is executed exclusively on a trusted thread, thereby preventing inconsistent PAC key usage. Since attackers cannot access the PAC module, they are hindered from bypassing our defenses.

For scalability, we encapsulate system calls for signature and authentication (\ssecref{subsec:pac_sign}) in the TEE OS, enabling resistance to new resource ID attacks by utilizing these interfaces with parameter variation. For compatibility, we implement \sysname in kernel space (i.e., Rich OS and TEE OS in EL1) without altering CAs and TAs. It maintains the existing invocation logic of TEE Client APIs and TEE Internal APIs. For security, we sign and authenticate only the resource IDs that identify TEE resources, resulting in fewer than a thousand lines of modifications to the TCB (\ssecref{subsec:tcb}). For efficiency, we introduce a minimal performance overhead (Table~\ref{tab:sumPerformance}) using a hardware-based authentication mechanism.

\section{Design} \label{sec:design}
In this section, we introduce the four main modules of \sysname: CA identity management (\ssecref{subsec:identity_management}) in the Rich OS, exception handling (\ssecref{subsec:exception}) and PAC signature and authentication (\ssecref{subsec:pac_sign}) in both the Rich OS and TEE OS, trusted thread management (\ssecref{subsec:thread_management}) in the TEE OS.

\subsection{CA Identity Management} \label{subsec:identity_management}

To address the potential security risk arising from the lack of randomness in PIDs used as identifiers for CA processes, we employ a 32-bit random number (\texttt{random$\_$id}) to represent the identity of CA processes instead. The \texttt{random$\_$id} is generated during the process initialization phase and is part of the process descriptor (\texttt{task$\_$struct}), ensuring its lifecycle is synchronized with the process. When CAs call TEE Client APIs to request TEE services, the Rich OS's TEE driver encapsulates the request for \texttt{SMC} instructions, places the CA process's \texttt{random$\_$id} in register x4, then forwards it to the TEE for further processing. A TA session will be created for each CA request, and we add the \texttt{random$\_$id} field in the TA session to store the identity of the requesting CA while initializing TA sessions, utilized for identity authentication.

\subsection{Exception Handling} \label{subsec:exception}

\sysname utilizes the Pointer Authentication Code (PAC) feature from the Armv8.3-A architecture (\ssecref{subsec:armPA}). To defend against brute-force attacks, the system handles exceptions differently between the Rich OS and the TEE OS. In the Rich OS (for CWE-639), a PAC validation failure results in the termination of the current CA process. Conversely, in the TEE OS (for CWE-862 and CWE-732), we introduce a \texttt{pac$\_$fail} flag into the TA session. If PAC validation fails, the \texttt{pac$\_$fail} flag is set, triggering an exception. During the exception handling, the system clears the current session and sends an error code to the TEE driver. The system then checks if the error code matches \texttt{TEE$\_$ERROR$\_$PAC$\_$FAIL} (a specific error code we defined). If it does, the current CA process is immediately terminated. Since OP-TEE lacks a recovery mechanism, direct termination of the process effectively prevents potential harm.

\begin{lstfloat}
\begin{lstlisting}[style=c-style, numbers=none, commentstyle=\color{blue}\itshape]
TEE_Result syscall_pacia (flag, data) 
{
    // Determine context based on flag
    if (flag == SESSION_PUBLIC)
        context = session->random_id;
    else
        context = (session->id << 32) + session->random_id;  

    // Execute PACIA signing
    asm("pacia %[reg], %[mod]" : [reg] "+r" (data) : [mod] "r" (context));
}

TEE_Result syscall_autia (flag, data) 
{
    // Determine context like syscall_pacia
    ......
    
    // Execute AUTIA verification
    asm("autia %[reg], %[mod]" : [reg] "+r" (data) : [mod] "r" (context)); 
    
    // Check PAC error bits and return error if set
    if (data & 0x6000000000000000) 
    {
        session->pac_fail = true;
        return TEE_ERROR_PAC_FAIL;
    }
}
\end{lstlisting}
\caption{The system calls used in TEE OS to sign and authenticate PAC.}
\label{lst:syscalls}
\end{lstfloat}

\subsection{PAC Signature and Authentication} \label{subsec:pac_sign} 

We employ PA instructions to sign the resource ID using user identity and authenticate the signature upon resource access, preventing SGVs arising from resource ID abuse. The PAC signature and authentication process is bifurcated into two parts (Fig.~\ref{fig:MaTEE}): the part in Rich OS addresses CWE-639 (\ssecref{subsec:auth_bypass}), while the part in TEE OS counters CWE-862 (\ssecref{subsec:missing_auth}) and CWE-732 (\ssecref{subsec:permission}).

In the Rich OS, PAC key management ensures that each user-level thread is associated with a unique key for PAC signatures, maintaining consistency throughout the thread's lifecycle. When a thread terminates, its key is replaced with a new one, ensuring randomness in the subsequent PAC generated, thereby mitigating the risk of replay attacks. We directly utilize PA instructions, employing \texttt{pacia} for signing and \texttt{autia} for authentication.

In the TEE OS, we introduce two system call functions to sign resource IDs, \texttt{syscall$\_$pacia()} and \texttt{syscall$\_$autia()}, detailed in Listing~\ref{lst:syscalls}. When the \texttt{SESSION$\_$PUBLIC} flag is enabled, the context for the PA instruction relies solely on the CA's process ID (i.e., \texttt{random$\_$id}), ensuring resource isolation among different processes. In contrast, when the flag is set to \texttt{SESSION$\_$PRIVATE}, the context integrates both the process ID and session ID. Since each thread within the same CA process is associated with a unique TA session for the TEE service, this approach achieves isolation among various CA threads.
If the \texttt{syscall$\_$autia()} verification fails and an error code is generated, the system will proceed with the exception handling procedure described in \ssecref{subsec:exception}.
Each request from the normal world is scheduled to run on a secure world trusted thread, which contains a static PAC key that does not change after the boot phase. Despite this, each CA process configuration in \ssecref{subsec:identity_management} is characterized by a distinctive, non-recurring \texttt{random$\_$id}. This setup renders replay attacks ineffective when setting \texttt{random$\_$id} as the PAC context in the \texttt{SESSION$\_$PUBLIC} mode. To further enhance security in the \texttt{SESSION$\_$PRIVATE} mode, we add a timestamp to each session ID.

\subsection{Trusted Thread Management} \label{subsec:thread_management}

The TEE OS assigns idle trusted threads to process requests from CAs. However, the system can dispatch calls from the same CA process to different trusted threads, potentially leading to key inconsistencies during signing and verification. To address this, we refined the trusted thread allocation by adding the \texttt{random$\_$id} field, which identifies the invoking CA process, to the trusted thread. We also tweaked the scheduling approach. When the TEE OS associates a CA process with a trusted thread, it first searches for threads with a matching \texttt{random$\_$id}. A match triggers the thread's state transition to active, enabling reuse. If no matches arise, the system commissions a fresh trusted thread for the CA process and records the associated \texttt{random$\_$id}. This enhancement ensures exclusive trusted threads for each CA process, resolving the key inconsistency challenge.

We further examine the impact of interruptions on trusted thread execution. Trusted thread execution can be interrupted by internal or external interrupts. For internal interrupts, the trusted thread resumes directly after handling the interrupt, without switching threads. In contrast, for external interrupts handled by the normal world, the trusted thread pauses. It resumes only after the normal world has addressed the interrupt or fulfilled the requested service, such as file access or retrieval of the current REE time. Our enhancements to trusted thread scheduling ensure that the same thread resumes execution after such external interruptions.

\section{Defend against SGVs} \label{sec:defend}

In this section, we present specific defense measures for Type-\RNum{2} SGVs, including preventing authentication bypass in the Rich OS (CWE-639, \ssecref{subsec:auth_bypass}), fixing missing authentication in the TEE OS (CWE-862, \ssecref{subsec:missing_auth}), and optimizing permission assignment in the TEE OS (CWE-732, \ssecref{subsec:permission}).

\subsection{Preventing Authentication Bypass} \label{subsec:auth_bypass}

\textbf{Protecting Shared Memory and TA Sessions.} 
We enhance the TEE driver to sign and authenticate session IDs and shared memory IDs using the PAC module (\ssecref{subsec:pac_sign}). This enhancement mitigates risks of authentication bypass, as attackers cannot access the PAC key to forge signatures. The signatures, embedded in the high bits of these IDs, require no additional storage space to be allocated.

For shared memory protection, we modify the TEE driver to employ the \texttt{pacia} instruction for signing shared memory IDs post-allocation and registration. Communication between CAs and TAs utilizes two types of parameters: data and memory reference. A memory reference parameter is comprised of a base address and an offset, and requires the memory to be pre-registered or allocated within a shared memory space. The TEE Client APIs then encapsulate and forward the CA's parameters to the TEE driver. When a parameter is identified as a memory reference by the TEE driver, we validate the shared memory ID's signature using the \texttt{autia} instruction. Successful verification allows us to strip the signature from the high-order bits seamlessly, ensuring that the parameter processing in the TEE OS remains unaffected. If the verification fails, the exception handling module (\ssecref{subsec:exception}) intervenes to resolve the issue.

For TA session protection, we modify the session creation module to sign the session IDs and validate them during their respective invocation functions (i.e., \texttt{tee$\_$ioctl$\_$invoke()}, \texttt{tee$\_$ioctl$\_$cancel()}, and \texttt{tee$\_$ioctl$\_$close$\_$session()}). The post-verification procedures for these session IDs are similar to those for the shared memory IDs.

\begin{table}[!t]
  \caption{Opaque handles and their purposes, as well as the count of TEE Internal APIs that \sysname modifies.}
  \label{tab:OpaqueHandles}
  \centering

  \begin{tabular}{l|l|c}
  \hline

  \hline
  \textbf{Opaque Handles} & \textbf{Objects handled} & \textbf{Counts} \\
  \hline

  \hline
    TASessionHandle  & sessions opened by a TA & 5\\
    \hline
    PropSetHandle  & property sets or enumerators & 13\\
    \hline
    ObjectHandle  & cryptographic or persistent objects & 29\\
    \hline
    ObjectEnumHandle  & persistent object enumerators & 6\\
    \hline
    OperationHandle  & cryptographic operations & 38\\
  \hline

  \hline
  \end{tabular}
\end{table}

\subsection{Fixing Missing Authentication} \label{subsec:missing_auth}

\textbf{Protecting Global Variables.}
We utilize LLVM to analyze the source code of TAs. If a TA is set to operate in single-instance mode and contains user-defined global variables, we halt its compilation and integration. This precaution is vital to prevent attackers from exploiting sensitive global variables through Type-\RNum{2} SGVs. As a best practice, developers are advised to use heap variables instead of global variables for data sharing within the TA. These heap addresses can either be attached to the TA's session context or associated with a global pointer. Each TA session receives a unique context pointer, facilitating state continuity across different commands within the same session and protecting against unauthorized access or modification of session-specific state information by other sessions.

For scenarios requiring data sharing across sessions, the variable is connected to a global pointer (\texttt{tee\_api\_instance\_data}), which is exclusively accessible through specific TEE Internal API interfaces, namely \texttt{SetInstanceData()} and \texttt{GetInstanceData()}. We refine the implementation of these interfaces: prior to linking the shared variable's address to the global pointer with \texttt{SetInstanceData()}, we initiate \texttt{syscall\_pacia()} to sign this address, activating the \texttt{SESSION\_PUBLIC} flag for sharing purposes. Conversely, when \texttt{GetInstanceData()} is utilized, we perform \texttt{syscall\_autia()} to verify the address and its high-bit signature, ensuring that only threads from the same CA process can access the shared variable.

\begin{lstfloat}
\begin{lstlisting}[style=c-style, numbers=none, commentstyle=\color{blue}\itshape]
// 1 Allocation APIs
TEE_Result CreatePersistentObject(..., ObjectHandle *object)
{
    // Set default flag to SESSION_PUBLIC
	uint32_t flag = SESSION_PUBLIC;

	if (object && *object == SESSION_PRIVATE)
		flag = SESSION_PRIVATE;

    // Allocate the opaque handle
    ......
    
    // Sign the opaque handle
    syscall_pacia(flag, *object);
}

// 2.1 Usage APIs -- Direct APIs
TEE_Result WriteObjectData(ObjectHandle object, ...)
{
    ObjectHandle obj = object;
    
	// Authenticate with SESSION_PUBLIC flag
    res = syscall_autia(SESSION_PUBLIC, object);
    
    // Try again using the SESSION_PRIVATE flag
	if (res == TEE_ERROR_PAC_FAIL) {
		object = obj;
		res = syscall_autia(SESSION_PRIVATE, object);
	}
 
    // Trigger a panic with the result error
	if (res != TEE_SUCCESS)
		TEE_Panic(res);

    ......
}

// 2.2 Usage APIs -- Wrapper APIs
TEE_Result __GP11_WriteObjectData(ObjectHandle object, ...)
{
    // Call the direct API 
	return WriteObjectData(object, ...);
}

// 2.3 Usage APIs -- Mixed APIs
void RestrictObjectUsage(ObjectHandle object, ...)
{
    ObjectHandle temp_obj = object;
    
    // Authenticate the handle like direct APIs
	check_object(&object);
 
    // Use the handle without PAC
	cryp_obj_get_info(object, ...);
 
    // Bring PAC back to the handle
	object = temp_obj;
 
    // Call the direct API like wrapper APIs
	RestrictObjectUsage1(object, ...);
}
\end{lstlisting}
\caption{Examples of allocation and usage APIs involving opaque handles.}
\label{lst:set_flag}
\end{lstfloat}

\textbf{Protecting TA Heap.} 
The heap address can be returned directly to CAs, serving as another alternative to global variable sharing. In OP-TEE, each data transfer from the REE to the TEE requires the data type to be specified. To safeguard heap variables, we introduce two new parameter types: \texttt{INVARIANT$\_$VALUE$\_$INPUT} (facilitating data flow from the REE to the TEE) and \texttt{INVARIANT$\_$VALUE$\_$OUTPUT} (facilitating data flow from the TEE to the REE). These types use PA to encapsulate heap variables. For \texttt{INVARIANT$\_$VALUE$\_$OUTPUT}, we generate the PAC using \texttt{syscall$\_$pacia()}; for \texttt{INVARIANT$\_$VALUE$\_$INPUT}, we verify the heap address using \texttt{syscall$\_$autia()} and remove the PAC before transmitting the parameters to the TEE. Additionally, we set a flag to switch the status, isolating or sharing heap variables among threads. Furthermore, upon receiving the data, the TA verifies that the data matches the specified type. Therefore, even if a malicious CA alters the type of a heap address—from \texttt{INVARIANT$\_$VALUE$\_$INPUT} or \texttt{INVARIANT$\_$VALUE$\_$OUTPUT} to another type that omits PAC authentication—it cannot circumvent the type verification implemented on the TEE side.

\subsection{Optimizing Permission Assignment} \label{subsec:permission}
\textbf{Protecting Opaque Handles.} 
An opaque handle serves as an abstract interface exposed to the TA by the TEE OS upon the creation of a sensitive resource object. Malicious CAs can compromise the sensitive resource using the forged opaque handles. This paper discusses five opaque handles, as shown in Table~\ref{tab:OpaqueHandles}, and modifies 91 TEE Internal APIs to protect these handles. This paper does not cover other opaque handles related to peripherals and events (e.g., \texttt{PeripheralHandle}), as these are currently not implemented in OP-TEE.

We implement the signing and authentication of opaque handles using previously encapsulated system calls (\ssecref{subsec:pac_sign}). We also identify all APIs that include opaque handles as parameters and categorize them into allocation and usage types (Listing~\ref{lst:set_flag}). For allocation APIs, when the TA requests TEE resources, we allocate and sign the opaque handles using \texttt{syscall$\_$pacia()}. To ensure compatibility and eliminate the need to rewrite TA code, we maintain the original API calling conventions. The signing flag is set to \texttt{SESSION$\_$PUBLIC} by default, switching to \texttt{SESSION$\_$PRIVATE} when resource isolation among threads is necessary. For usage APIs, we authenticate opaque handles with \texttt{syscall$\_$autia()} prior to use. To avoid redundant verifications due to mutual API invocations, we divide these APIs into three categories (Listing~\ref{lst:set_flag}):

\begin{enumerate}
    \item \textbf{Direct APIs}: These APIs function independently, without calling other usage APIs. We use \texttt{syscall$\_$autia()} to validate their handles. Once the verification is successful, the PAC in the higher order bits is removed.
    \item \textbf{Wrapper APIs}: These are essentially wrappers around direct APIs. We pass the handles with the PAC still attached directly through these APIs.
    \item \textbf{Mixed APIs}: These APIs involve calling direct APIs while also managing opaque handles internally. We first authenticate and then remove the PAC for internal use of the handles. However, when passing handles to direct APIs, the validated handles are provided with the PAC intact.
\end{enumerate}

\begin{figure}[!t]
  \centering
  \includegraphics[width=3.5 in]{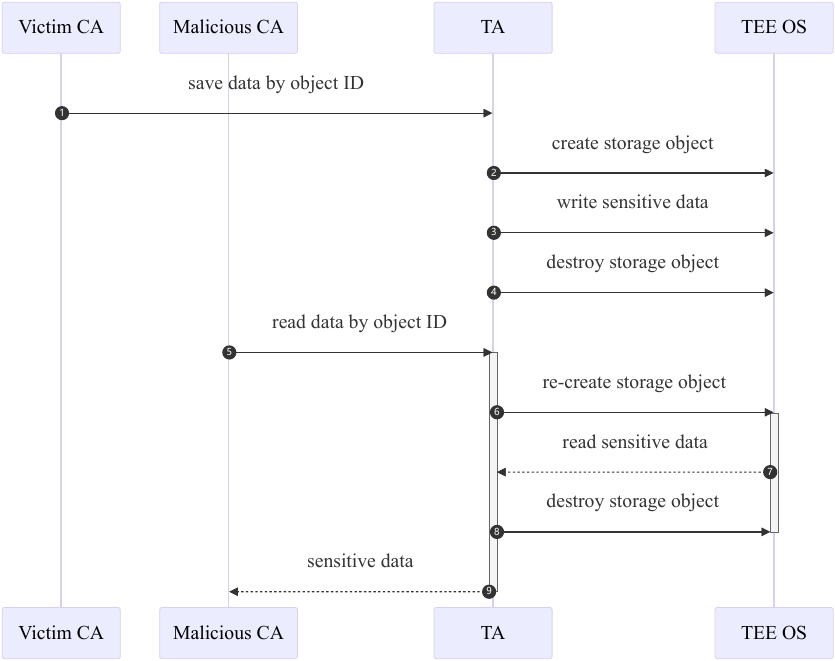}
  \caption{Persistent storage leakage.}
  \label{fig:storageHijack}
\end{figure}

\textbf{Protecting Persistent Storage.} 
Despite safeguards on opaque handles for persistent storage (i.e., \texttt{ObjectHandle} and \texttt{ObjectEnumHandle}), adversaries can still create new signed opaque handles using object IDs. This allows them to circumvent PAC authentication and gain unauthorized access to persistent storage, as depicted in Fig.~\ref{fig:storageHijack}. Specifically, the victim CA uses the object ID to instantiate a storage object with an opaque handle signed using the victim CA's identity, and then stores sensitive data in the persistent storage managed by this object. Upon termination of the writing process, the storage object is closed and immediately deconstructed. Adversaries can exploit the same object ID to re-create a storage object to access the same persistent storage, but with a new opaque handle signed using the malicious CA's identity.

To mitigate this vulnerability, we modify the procedure that closes the storage object to reset only the flags and reference count, thereby preserving the storage object. When an adversary attempts to access persistent storage, the adversary will use an opaque handle that has been signed with the identity of the victim CA, rather than that of the malicious CA. Since the signature on this opaque handle cannot pass the PAC verification, \sysname can effectively defend against this type of attack.

\begin{table*}[!t]
  \caption{Analysis of mitigation for Type-\RNum{2} SGVs. In TA, $M$ stands for Multi-instance, and $S$ stands for Single-instance. Symbol \CIRCLE~~indicates that this setup can resist SGVs between CA threads and processes, symbol \LEFTcircle~~indicates that this setup can only resist SGVs between CA processes, and symbol \Circle~~indicates that this setup cannot resist SGVs.}
  \label{tab:mitigation}
  \centering

  \begin{tabular}{l|c|c|c|c|c}
  \hline

  \hline
  \multirow{2}{*}{\textbf{Type-\RNum{2} SGVs}} & \multirow{2}{*}{\textbf{Exploited Resource Identifier}} & \textbf{Victim Resource} & \multirow{2}{*}{\textbf{TA}} & \textbf{\sysname} & \multirow{2}{*}{\textbf{\sysname}}\\
   & & \textbf{Manager \& Attack Type} & & \textbf{Disabled} & \\
  \hline

  \hline
    Hijack shared memory & Shared memory ID & \multirow{2}{*}{Rich OS (CWE-639)} & $M\&S$  & \LEFTcircle & \CIRCLE\\
  \cline{1-2} \cline{4-6} 
    Hijack TA sessions & Session ID & & $M\&S$  & \LEFTcircle & \CIRCLE\\
  \hline
    Compromise global variables & Global variables & \multirow{2}{*}{TA (CWE-862)} & $S$  & \Circle & \CIRCLE\\
  \cline{1-2} \cline{4-6} 
    Compromise TA heap & Heap address & & $S$  & \Circle & \CIRCLE\\
  \hline
    Hijack persistent storage & Storage ID & \multirow{6}{*}{TEE OS (CWE-732)} & $M\&S$  & \Circle & \CIRCLE\\
  \cline{1-2} \cline{4-6} 
    \multirow{5}{*}{Compromise by opaque handles} & ObjectHandle & & $M\&S$  & \Circle & \CIRCLE\\
    \cline{2-2} \cline{4-6}
     & TASessionHandle &  & $M\&S$  & \Circle & \CIRCLE\\
    \cline{2-2} \cline{4-6}
     & PropSetHandle &  & $M\&S$  & \Circle & \CIRCLE\\
    \cline{2-2} \cline{4-6}
     & ObjectEnumHandle &  & $M\&S$  & \Circle & \CIRCLE\\
    \cline{2-2} \cline{4-6}
     & OperationHandle &  & $M\&S$  & \Circle & \CIRCLE\\
  \hline
  
  \hline
  \end{tabular}
\end{table*}

\section{Implementation} \label{sec:prototype}
In this section, we develop two prototypes: a functional prototype validated for security and compatibility on the Arm FVP simulator, and a performance prototype evaluated on the Rock Pi 4B development board.

\textbf{Functional Prototype.} 
We implement \sysname on the open-source and popular OP-TEE platform. The functional prototype is developed on the Arm Fixed Virtual Platform (FVP), which supports the Armv8.3-A PA feature. Utilizing the PA in \sysname necessitates enabling both runtime and compilation support. Activating runtime support involves turning on the FVP's \texttt{has$\_$arm$\_$v8-3} boot parameter, along with Trusted Firmware-A's \texttt{ENABLE$\_$PAUTH} and \texttt{CTX$\_$INCLUDE$\_$PAUTH$\_$REGS}, and TEE OS's \texttt{CFG$\_$CORE$\_$PAUTH}. For compilation support, the \texttt{-march=armv8.3-a} option should be added to the compile scripts of Linux and OP-TEE OS.

\textbf{Performance Prototype.}
We port the performance prototype to the Rock Pi 4B. This board features a six-core 64-bit Rockchip RK3399 SoC, including two Arm Cortex-A72 cores (up to 1.8GHz) and four Cortex-A53 cores (up to 1.4GHz), and is equipped with 4GB LPDDR4 RAM and 32GB eMMC storage. It runs Linux 6.2.0-rc3 in the normal world and OP-TEE OS 3.20.0 in the secure world. Due to the absence of Armv8.3-A PA support on the Rock Pi 4B development board, we simulate the PA instruction through software, following approaches used in previous work~\cite{farkhani2021ptauth},~\cite{schilling2022fipac}. We substitute all PA instructions with four consecutive XOR operations. While this approach may not offer security comparable to the original hardware implementation, it suffices for performance evaluation. The actual performance is likely better than our simulation since hardware PAC offers greater execution efficiency. Additionally, while initiatives like PACMem~\cite{li2022pacmem} have successfully implemented PA instructions in user space on the Apple M1 Mac mini, we cannot deploy \sysname on it due to Apple's proprietary SoC and TEE (Secure Enclave~\cite{Apple2010Platform}).

\section{Security Evaluation} \label{sec:security_evaluation}

In this section, we first measure the TCB size of \sysname (\ssecref{subsec:tcb}). We then test \sysname's defense against SGVs (\ssecref{subsec:mitigation}). Finally, we analyze the attacks permitted under our threat model and explain how \sysname prevents them (\ssecref{subsec:secutity_analysis}).

\subsection{TCB} \label{subsec:tcb}

\begin{table}[!t]
  \caption{Total modifications to the TCB required to implement \sysname, measured in LoCs.}
  \label{tab:LOC}
  \centering

  \begin{tabular}{l|ccc}
  \hline

  \hline
  \textbf{Component} & \textbf{Added} & \textbf{Modified} & \textbf{Total} \\
  \hline

  \hline
    TEE Client Library & 15 & 9 & 24\\
  \hline
    Linux & 58 & 11 & 69\\
  \hline
    Trusted Firmware-A & - & 2 & 2\\
  \hline
    OP-TEE OS & 772 & 68 & 840\\
  \hline
    Static Analysis Module & 47 & 2 & 49\\
  \hline
    \textbf{All} & 892 & 92 & 984\\
  \hline

  \hline
  \end{tabular}
\end{table}

We introduce 984 lines of code (LoCs) of modifications (Table~\ref{tab:LOC}) to the TCB without altering the architecture of Linux or OP-TEE OS. We modify the TEE Client Library and add new parameter types to protect the integrity of the TA heap address (\ssecref{subsec:missing_auth}). Additionally, we modify Linux to introduce the CA identity management module (\ssecref{subsec:identity_management}) and enhance OP-TEE OS with the trusted thread management module (\ssecref{subsec:thread_management}). Moreover, we implement the PAC generation and authentication module (\ssecref{subsec:pac_sign}) and the exception handling module (\ssecref{subsec:exception}) in both Linux and OP-TEE OS. Furthermore, Trusted Firmware-A is modified to enable PAC support (\ssecref{sec:prototype}), and a static analysis module is incorporated (\ssecref{subsec:missing_auth}) to prevent global variable hijacking.

\subsection{SGVs Mitigation Analysis} \label{subsec:mitigation}
As noted earlier, OP-TEE has robust defenses against Type-\RNum{1} SGVs. Since \sysname is built on top of OP-TEE, it naturally inherits these protective features.
Given the absence of CVEs related to Type-\RNum{2} SGVs (CVE-2016-5349 targets Type-\RNum{1} SGVs), we meticulously developed and open-sourced specific test cases\footnotemark\footnotetext{\url{https://github.com/erhade/MaTEE/blob/master/src/optee_examples/semantic_victim}} targeting six types of Type-\RNum{2} SGVs. These test scenarios encompass established vulnerabilities like \textsc{Boomerang}~\cite{machiry2017boomerang} (exploiting shared memory IDs) and HPE~\cite{suciu2020horizontal} (exploiting session IDs, global variables, and storage IDs), as well as new potential threats we have identified (exploiting heap address and five opaque handles detailed in Table~\ref{tab:OpaqueHandles}). We establish a pair of CA and TA, where the victim CA requests exploitable shared resources such as persistent storage from the TA. Subsequently, we introduce a malicious CA that connects to the same TA (using the same TA \texttt{UUID}) to illicitly access sensitive data, such as cryptographic keys, by exploiting the same resource ID. The detailed attack scenarios are as follows:

\textbf{Shared Memory Hijacking}: The victim CA registers and initializes shared memory, which the malicious CA exploits by using the same memory ID to alter the stored data.

\textbf{TA Session Hijacking}: The malicious CA takes over the session ID used by the victim CA, gaining control of the TA session and prematurely ending it.

\textbf{Global Variable Compromise}: The victim CA requests the TA to allocate heap memory and link it to a global pointer managed by TEE Internal APIs. After sensitive data is stored in this memory, the malicious CA initiates a new TA session to access this data using the global pointer.

\textbf{TA Heap Compromise}: Similar to the attack on global variables, the heap memory allocated by the TA for the victim CA can be compromised. Instead of being linked to a global pointer, the address is directly returned to the CA, allowing the malicious CA to access and extract sensitive data from the heap.

\textbf{Persistent Storage Hijacking}: The victim CA uses a specific storage ID to save sensitive data in the TA-managed persistent storage. The malicious CA exploits this same storage ID to retrieve the stored data.

\textbf{Opaque Handle Exploitation}: Using the cryptographic object handle (i.e., \texttt{ObjectHandle}) as an example, the victim CA requests the TA to generate a key and return an \texttt{ObjectHandle}. The malicious CA uses this handle to turn the TA into a cryptographic oracle, enabling unauthorized decryption or encryption of data.

As detailed in Table~\ref{tab:mitigation}, we evaluate the efficacy of CA inter-process and inter-thread isolation across single and multiple TA instances against the described attacks. While the OP-TEE driver monitors session and shared memory IDs for each CA process, providing partial protection against Type-\RNum{2} SGVs between CA processes, it is still susceptible to four other types of Type-\RNum{2} SGVs. Our analysis robustly supports the conclusion that \sysname provides effective defense against all six types of Type-\RNum{2} SGVs.

\begin{table}[!t]
  \caption{Summary of performance evaluation.}
  \label{tab:sumPerformance}
  \centering

  \begin{tabular}{l|c|c}
  \hline

  \hline
  \textbf{Tests} & \textbf{Test Cases}& \textbf{Overhead}\\
  \hline
    TEE Client APIs & 9 & 1.69$\%$\\
  \hline
    \texttt{xtest} regressions & 10 & 1.39$\%$\\
  \hline
    \texttt{xtest} benchmarks & 28 & 4.19$\%$\\
  \hline
    AVB & 6 & -2.67$\%$\\
  \hline
    Trusted keys & 3 & 0.50$\%$\\
  \hline
    DarkneTZ & 10 & 1.27$\%$\\
  \hline
    All tests & 66 & 2.19$\%$\\
  \hline

  \hline
  \end{tabular}
\end{table}

\subsection{Security Analysis} \label{subsec:secutity_analysis}

\textbf{Resource ID Impersonation.} 
Our threat model permits the adversary to obtain resource IDs of other CAs (high bits with PAC). However, the PAC verification process (\ssecref{subsec:pac_sign}) demands the CA's identity and the corresponding TA session ID as the PA context. As a result, this verification process prevents the adversary from directly exploiting the acquired resource IDs to gain unauthorized access to sensitive data of other CAs.

\textbf{PA instruction Reuse.}
In our threat model, adversaries face limitations in controlling privileged software, which hinders their ability to execute PA instructions at the privileged level. Although they might attempt to create signatures by invoking PA instructions at the user level, these signatures will fail the verification process (\ssecref{subsec:pac_sign}) due to the distinct PAC keys used at the user and privileged levels.

\textbf{Brute-Force Attacks.}
Our threat model considers the possibility of adversaries attempting to forge PAC through brute-force attacks. We have implemented safeguards to counter such threats. First, the PAC length in the secure world is set to 31 bits (\ssecref{subsec:armPA}), making brute-forcing exceedingly costly and time-consuming. Second, our system incorporates an exception-handling module (\ssecref{subsec:exception}) that promptly terminates any malicious adversary making incorrect guesses. These countermeasures ensure that adversaries can make only one guess before being terminated, effectively defending against brute-force attacks.

\textbf{Replay Attacks.} 
In our threat model, adversaries can replay pre-requested signed resource IDs. For PAC authentication in the Rich OS (\ssecref{subsec:auth_bypass}), replay attacks are ineffective as distinct PAC keys are used for adversary threads. For PAC authentication in the TEE OS (\ssecref{subsec:missing_auth} and \ssecref{subsec:permission}), we ensure that each PAC signature is created with a unique context, specific to each CA process and TA session.

\section{Performance Evaluation} \label{sec:performance_evaluation}
In this section, we employ the original OP-TEE as a baseline to evaluate the performance overhead of \sysname. Our performance evaluation covers three vectors. First, we evaluate the overhead of TEE Client APIs to demonstrate \sysname's influence on the REE side (\ssecref{subsec:ree_side}). Second, we assess \sysname's compatibility and performance using the official OP-TEE \texttt{xtest}~\cite{OPTEExtest} (\ssecref{subsec:microbenchmarks}). Third, we port \texttt{AVB}~\cite{Wiklander2018avb}, \texttt{Trusted keys}~\cite{Garg2020trustedkeys}, and \texttt{DarkneTZ}~\cite{mo2020darknetz} into \sysname to measure the performance impact on real-world applications (\ssecref{subsec:applications}). We conduct each test case over 100 times to calculate the average overhead. After each test, we flush the cache to avoid the effects of hot caching on runtime. A series of 66 sub-tests are carried out, indicating an average performance overhead of 2.19$\%$ for \sysname, as depicted in Table~\ref{tab:sumPerformance}.

\begin{figure}[!t]
  \centering
  \includegraphics[width=3 in]{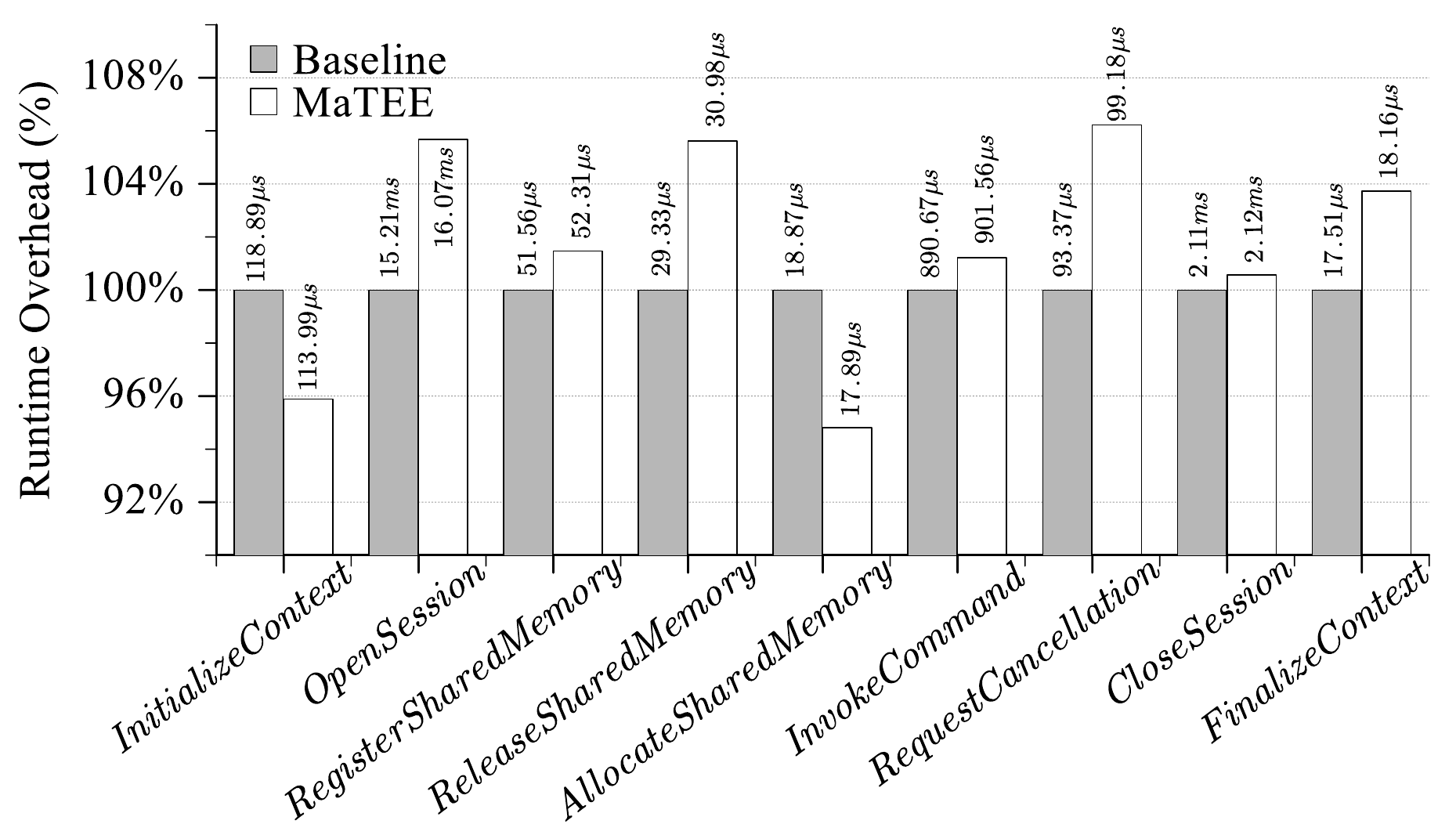}
  \caption{The runtime overhead of TEE Client APIs.}
  \label{fig:ClientAPI}
\end{figure}

\subsection{Impact on the REE Side} \label{subsec:ree_side}
We compare the TEE client API execution time between \sysname and the baseline, as depicted in Fig.~\ref{fig:ClientAPI}. Our findings demonstrate that \sysname has minimal impact on REE performance, with an average overhead of only 1.69$\%$. Except for the \texttt{OpenSession()} operation, the performance overhead remains at the microsecond level. The \texttt{OpenSession()} operation incurs a performance cost of 864.153 $\mathrm{\mu}$s. The overhead for all operations does not exceed 7$\%$, with the highest performance cost observed for the \texttt{RequestCancellation()} operation at 6.23$\%$.

\begin{table*}[!ht]
  \caption{The runtime overhead of Baseline (Base) and \sysname in three scenarios: Read, Rewrite, and Write, under different data sizes.}
  \label{tab:benchmark}
  \centering

  \begin{tabular}{c|c|c|c|c|c|c}
  \hline

  \hline
  \textbf{Data Size (B)} & \textbf{Base-Read (s)} & \textbf{\sysname-Read (s)} & \textbf{Base-Rewrite (s)} & \textbf{\sysname-Rewrite (s)} & \textbf{Base-Write (s)} & \textbf{\sysname-Write (s)}\\
  \hline

  \hline
  256 & 0.00012 & 0.00014 & 0.00661 & 0.00712 & 0.00584 & 0.00579\\
  \hline
  512 & 0.00017 & 0.00019 & 0.00711 & 0.00752 & 0.00576 & 0.00581\\
  \hline
  1024 & 0.00024 & 0.00015 & 0.00779 & 0.00768 & 0.00635 & 0.00644\\
  \hline
  2048 & 0.0002 & 0.00027 & 0.01432 & 0.01458 & 0.01314 & 0.01406\\
  \hline
  4096 & 0.00052 & 0.00065 & 0.03252 & 0.03275 & 0.02876 & 0.02872\\
  \hline
  16384 & 0.00189 & 0.00189 & 0.14612 & 0.14475 & 0.1397 & 0.14099\\
  \hline
  524288 & 0.13012 & 0.14442 & 6.45046 & 6.45716 & 6.2511 & 6.32473\\
  \hline
  1048576 & 0.313 & 0.32087 &13.66065 & 13.80069 & 13.39987 & 13.44183\\
  \hline

  \hline
  \end{tabular}
\end{table*}

\subsection{Microbenchmarks} \label{subsec:microbenchmarks}
\begin{figure}[!t]
  \centering
  \includegraphics[width=3 in]{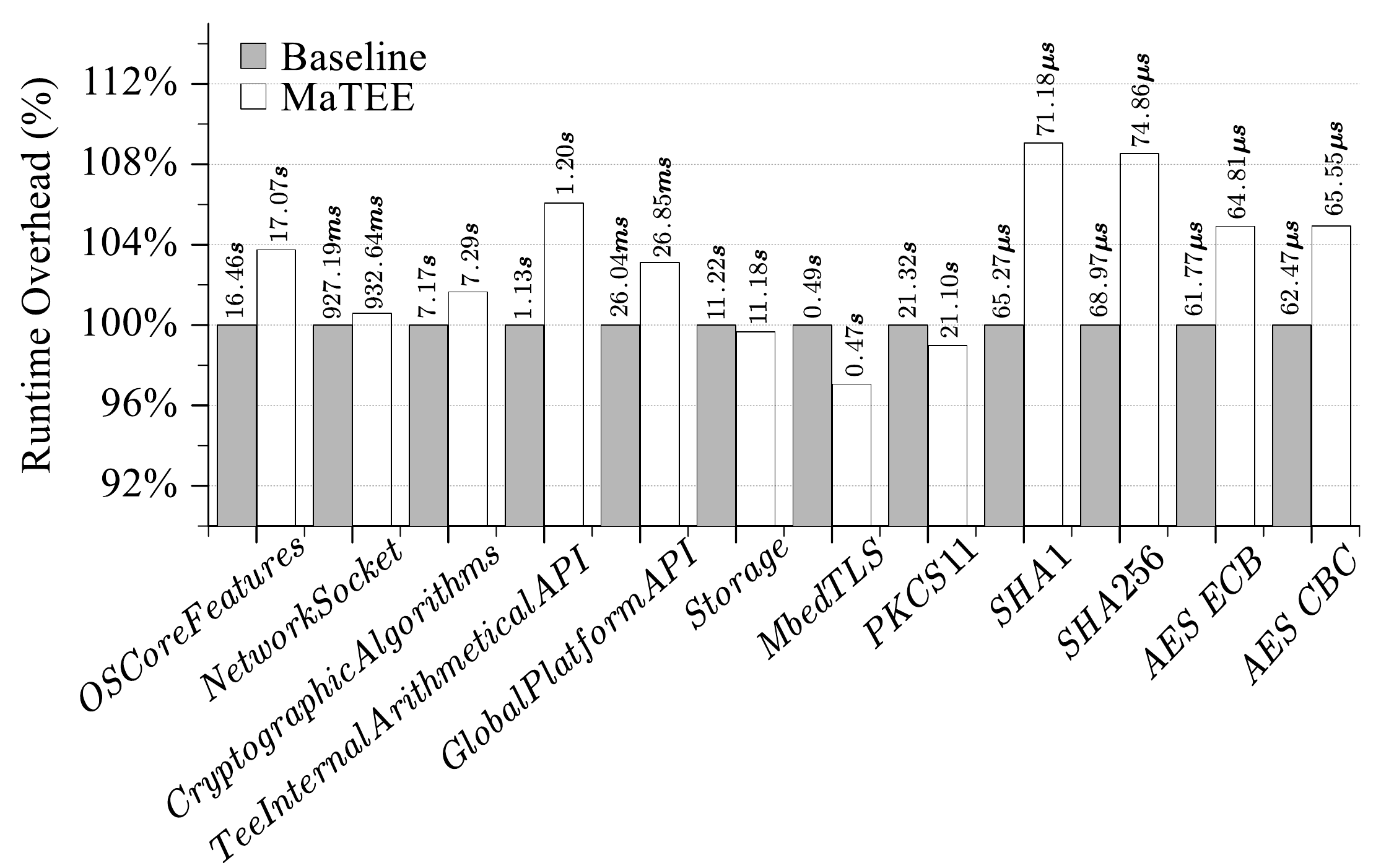}
  \caption{The runtime overhead of 8 sets of regression tests and 4 sets of benchmark tests for SHA1, SHA256, AES-ECB, and AES-CBC.}
  \label{fig:xtest}
\end{figure}

\sysname passes all 137 test cases (including 31,072 subtests) from the \texttt{xtest} regression test suite. Zero regression test errors suggest that \sysname is fully compatible with other system components and can be deployed directly into a production environment. We categorize the 137 tests into eight groups based on their functionality. For instance, the \texttt{OS Core Features} group includes 36 test cases, such as those for shared memory. We then compare each group's overhead with the baseline and the results are shown in the left side of Fig.~\ref{fig:xtest}. These eight groups of regression tests generate a total overhead of 1.36$\%$. The \texttt{TEE Internal Arithmetical API} is most affected, presenting an overhead 6.07$\%$ higher than the baseline, equating to an actual difference of 68.693 ms. The \texttt{Storage}, \texttt{Mbed TLS}, and \texttt{PKCS11} groups exhibit lower overheads than the baseline, which can be attributed to our modifications to trusted thread scheduling. We effectively improve performance by binding calls from the same CA to execute within a single trusted thread.

We implement two additional regression tests to evaluate the overhead of CA inter-thread memory sharing (\ssecref{subsec:missing_auth}). The first test uses the native \texttt{Set/GetInstanceData()} interface as the baseline and compares it with the PAC-modified interface to evaluate the overhead of memory sharing through global variables, showing an overhead of 0.19\%. The second test evaluates the overhead of memory sharing through the heap, using the \texttt{VALUE\_OUTPUT} type to pass heap addresses as the baseline and comparing it with the \texttt{INVARIANT\_VALUE\_OUTPUT} type (which undergoes PAC signature verification) to pass heap addresses, showing an overhead of 2.81\%. Overall, \texttt{xtext} regressions consist of a total of 10 test groups, with an average overhead of 1.39\%.

We deploy 28 \texttt{xtest} benchmark tests to further explore performance overhead of \sysname. These benchmark tests result in a total runtime overhead of 4.19$\%$. The performance outcomes for \texttt{SHA} and \texttt{AES} tests are illustrated on the right side of Fig.~\ref{fig:xtest}, showcasing a performance overhead of approximately 9$\%$ for \texttt{SHA} and around 5$\%$ for \texttt{AES}. Additionally, Table~\ref{tab:benchmark} presents a concise performance comparison between \sysname and the baseline for reading, writing, and rewriting scenarios. \sysname introduces overhead in all of these scenarios, with 8.05$\%$ overhead for reads, 1.85$\%$ for rewrites, and 1.34$\%$ for writes. 

\begin{figure}[!t]
  \centering
  \includegraphics[width=3 in]{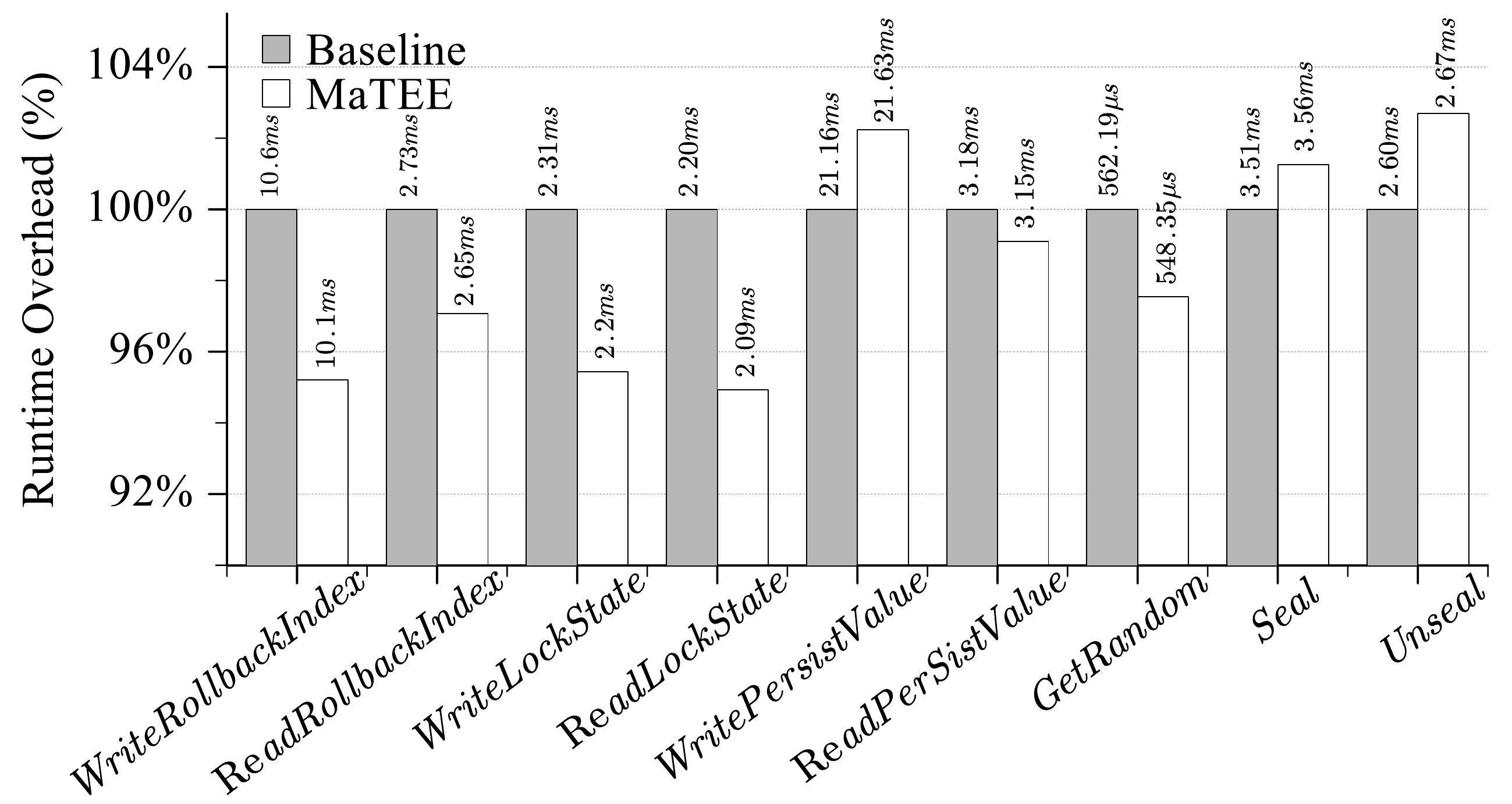}
  \caption{The runtime overhead of AVB and Trusted keys.}
  \label{fig:AVBAndTKeys}
\end{figure}

\subsection{Real-world Applications} \label{subsec:applications}

\texttt{Android Verified Boot (AVB)} is a security feature that ensures software integrity on devices by verifying signatures at each boot stage, using the TEE for secure storage. \texttt{Trusted keys} leverage the TEE for sealing and unsealing keys of the normal world. We modify OP-TEE's versions of both \texttt{AVB}~\cite{Wiklander2018avb} and \texttt{Trusted keys}~\cite{Garg2020trustedkeys} for our testing purposes. Our findings for \texttt{AVB}, illustrated in six test cases on the left side of Fig.~\ref{fig:AVBAndTKeys}, align with the \texttt{storage} test group in the \texttt{xtest} regression (\ssecref{subsec:microbenchmarks}) and show an average time overhead reduction of 2.67\%. For \texttt{Trusted keys}, depicted by three test cases on the right side of Fig.~\ref{fig:AVBAndTKeys}, the \texttt{GetRandom API} achieves a 2.46\% reduction in runtime. Nonetheless, the \texttt{Seal} and \texttt{Unseal} operations exhibit increased performance overhead compared to the baseline, leading to an overall 0.50\% overhead for \texttt{Trusted keys}.

\begin{figure}[!t]
  \centering
  \includegraphics[width=3.3 in]{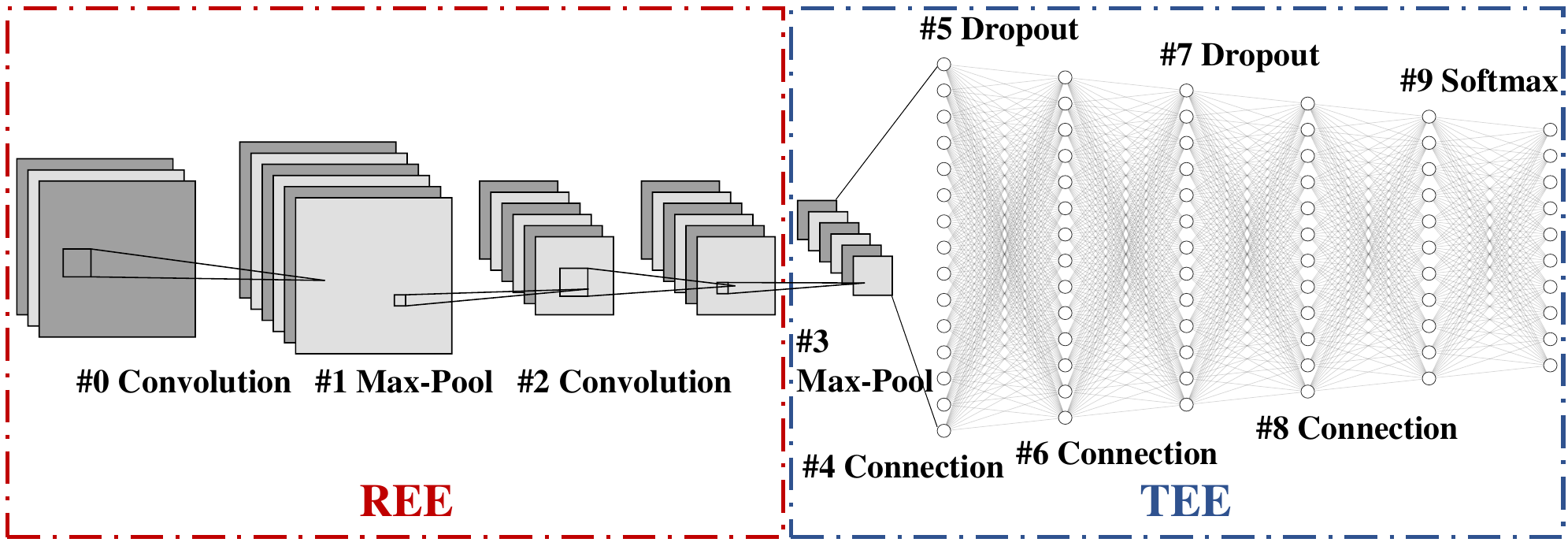}
  \caption{The structure of DarkneTZ. The first three layers run in the REE, with the subsequent seven layers operating in the TEE.}
  \label{fig:DNN}
\end{figure}

\begin{figure}[!t]
  \centering
  \includegraphics[width=3 in]{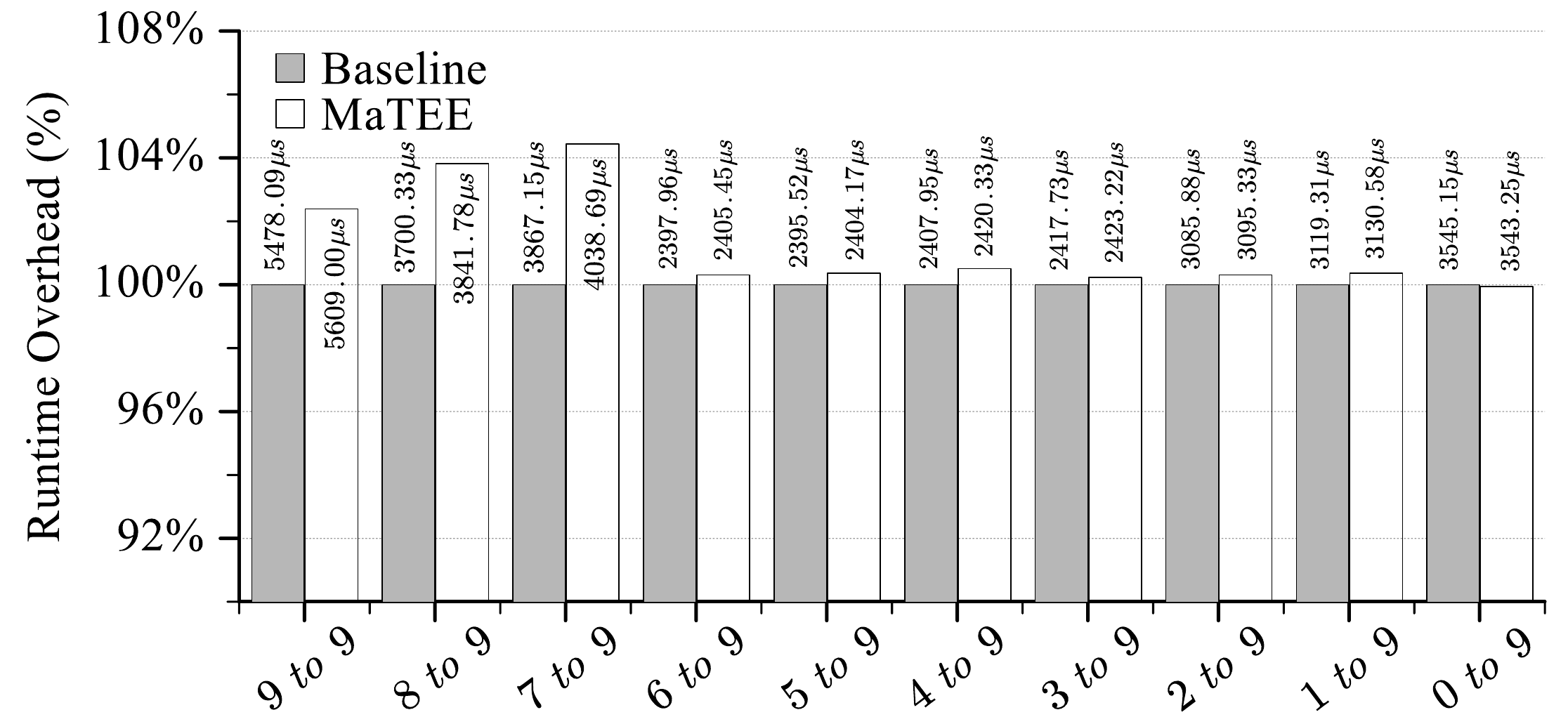}
  \caption{The runtime overhead of DarkneTZ during inference, where "m to n" represents the mth to nth layers running on the TEE side, while the other layers operate on the REE side.}
  \label{fig:DarkneTZ}
\end{figure}

\texttt{DarkneTZ}~\cite{mo2020darknetz} migrates a 10-layer deep neural network (DNN) used for image classification tasks into TrustZone. Its unique functionality enables specific sensitive layers to run within the TEE. In contrast, the remaining layers operate in the REE (Fig.~\ref{fig:DNN}). The network comprises convolutional and pooling layers for feature extraction, connection and dropout layers for regularization, and a softmax layer for classification, with a cost layer for error calculation. In our experiments (Fig.~\ref{fig:DarkneTZ}), we place a specific number of layers in the TEE and compare the performance overhead of \sysname against the baseline for each case. When all ten layers run in the TEE, \sysname shows a 0.05$\%$ reduction in execution time compared to the baseline. However, in other configurations, \sysname exhibits performance overhead exceeding the baseline. When only the dropout, connection, and softmax layers are in the TEE, \sysname has the highest performance overhead of 4.44$\%$. The overall average overhead is 1.27$\%$.

\section{Discussion} \label{sec:discussion}

\textbf{Long-Term Storage Protection.}
Currently, \sysname only supports runtime storage protection. If a CA maintains a confidential file and fails to remove it upon process termination, freshly launched CA processes cannot access it due to changes in the \texttt{random$\_$id}. If the TEE OS restarts, it eradicates all previously managed metadata linked to storage, making the file accessible to any CA that possesses the storage ID. Users and TAs should establish an identity authentication mechanism to safeguard long-term storage. This mechanism necessitates the persistent storage of user identities, rather than CA process IDs, in the metadata.

\textbf{Resource Sharing.} 
\sysname facilitates resource sharing across threads within the TEE OS, including the TA heap, opaque handles, and persistent storage. In contrast, resource sharing in the Rich OS is not currently supported. Nevertheless, altering the PAC key management in the Rich OS might enable this system to dynamically synchronize PAC keys between different threads based on their sharing needs.

\textbf{Backward Compatibility.} The full implementation of \sysname requires PAC support, which means it does not support devices older than Armv8.3-A. As demonstrated by xMP~\cite{proskurin2020xmp}, software-based schemes that ensure pointer integrity achieve backward compatibility and provide protection across platforms, including Intel and AMD.

\textbf{Availability.} Our exception-handling module (\ssecref{subsec:exception}) swiftly detects and terminates the malicious CA process to mitigate brute-force attacks. However, this immediate response may inadvertently enable the malicious CA to perform a denial-of-service attack. A potential solution is to cap the number of verification attempts for a given process within a set timeframe. Upon reaching this threshold, the system can introduce delays in verification or temporarily suspend the verification requests for the process.

\section{Related Work} \label{sec:related_work}
\textbf{Capabilities-Based Systems}. The concept of capabilities as a means to define and enforce fine-grained, principle-of-least-privilege security has existed for several decades.
CHERI~\cite{watson2015cheri},~\cite{joannou2017efficient},~\cite{woodruff2014cheri} extends conventional RISC architectures by introducing a rich set of capability primitives to the hardware. This hybrid capability-system architecture combines the performance and ubiquity of traditional CPUs with the robustness and security guarantees that come with pure capability systems. Similar to CHERI, \sysname employs hardware-based pointer authentication to differentiate between user capabilities and to enforce the corresponding TEE resource access isolation policies.

\textbf{PAC-Assisted Systems}. The capability-based \sysname utilizes PAC, which originates from the Arm AArch64 architecture, to provide an unforgable token. AOS-RISC-V~\cite{kimrisc} and \textsc{RetTag}~\cite{wang2022rettag} support PAC on the RISC-V platform, implemented on the FPGA. Several studies~\cite{schilling2022fipac},~\cite{liljestrand2019pac},~\cite{serra2022pac},~\cite{ismail2022tightly},~\cite{liljestrand2021pacstack} utilize Arm PAC to enhance Control-Flow Integrity (CFI) for user-level software. Others, such as Camouflage~\cite{denis2020camouflage}, PAL~\cite{yoo2022kernel}, and PATTER~\cite{yang2019arm}, employ PAC to enhance the kernel's CFI. HAKC~\cite{mckee2022preventing} applies PAC and MTE to the kernel module isolation beyond CFI. Furthermore, several works deploy PAC for memory safety. PTAuth~\cite{farkhani2021ptauth} is a runtime protection scheme for heap-based temporal safety. AOS~\cite{kim2020hardware} ensures heap spatial and temporal safety. PACMem~\cite{li2022pacmem} can provide complete memory safety, protecting a program's heap and stack from spatial and temporal memory corruptions. Unlike user-space or kernel-space protections, \sysname is the first scheme utilizing PAC to enhance the security of TEEs.

\textbf{TrustZone Vulnerabilities}.
SGVs~\cite{machiry2017boomerang},~\cite{suciu2020horizontal} are a type of confused deputy attack in TrustZone, exploiting the privileges of the TEE to launch attacks on the REE side. Over the years, researchers have discovered multiple potential attack vectors against TrustZone. Software vulnerabilities in TrustZone manifest in components such as the TEE OS~\cite{shen2015exploiting},~\cite{Beniamini2016CVE}, TAs~\cite{rosenberg2014reflections},~\cite{berard2018kinibi},~\cite{Komaromy2018Unbox}, and bootloaders~\cite{rosenberg2013unlocking}, all offering potential exploit avenues. In addition to software implementation vulnerabilities, TrustZone is also affected by side-channel attacks. Noteworthy among these are hardware threats like CLKscrew~\cite{tang2017clkscrew} and VoltJockey~\cite{qiu2019voltjockey}, which predominantly exploit voltage and power modulations to introduce faults or steal sensitive data from the secure world. The Rowhammer attack~\cite{van2016drammer}, another threat, compromises DRAM by instigating bit flips in memory units.

\textbf{TEE Privilege Reduction Techniques}. 
Several works reduce the privileges of the TEE by constructing new isolated regions in the secure world. \textsc{ReZone}~\cite{cerdeira2022rezone} uses commercial off-the-shelf (COTS) platforms to divide the TEE OS into different zones. TwinVisor~\cite{li2021twinvisor} leverages Arm S-EL2 to offer confidential virtual machines (VMs), similar to Hafnium~\cite{2023hafnium}. Other works establish enclaves in the normal world specifically to avoid expanding the Trusted Computing Base (TCB) of the secure world. TrustICE~\cite{sun2015trustice}, CacheIEE~\cite{wang2023cacheiee}, and Ginseng~\cite{yun2019ginseng} create Isolated Execution Environments (IEEs) to protect security-critical applications. Sanctuary~\cite{brasser2019sanctuary} enables an enclave to utilize a physical core exclusively, achieving memory isolation by modifying the TZASC configuration. In contrast, vTZ~\cite{hua2017vtz}, OSP~\cite{cho2016hardware}, and PrivateZone~\cite{jang2016privatezone} implement isolation by leveraging the virtualization extensions available at the normal world's EL2. However, unlike MATEE, these systems do not explicitly defend against SGVs.

\section{Conclusion}

In this paper, we introduce \sysname, a defense mechanism designed to mitigate Type-\RNum{2} SGVs. We address well-documented vulnerabilities like \textsc{Boomerang} and HPE, along with our discoveries related to the TA heap and opaque handle compromises. By leveraging Arm PA, we ensure resource isolation among CA processes and threads through authenticated resource IDs. Additionally, we present a proof of concept (POC) specific to Type-\RNum{2} SGVs. Our comprehensive evaluation validates \sysname's capability to counter all six recognized types of Type-\RNum{2} SGVs. Performance evaluations using microbenchmarks and real-world applications, such as \texttt{DarkneTZ}, demonstrate that \sysname incurs a minimum performance overhead of only 2.19\%.

\bibliographystyle{IEEEtran}
\bibliography{reference}

@inproceedings{mcgillion2015open,
  title={Open-TEE--an open virtual trusted execution environment},
  author={McGillion, Brian and Dettenborn, Tanel and Nyman, Thomas and Asokan, N},
  booktitle={2015 IEEE Trustcom/BigDataSE/ISPA},
  volume={1},
  pages={400--407},
  year={2015},
  organization={IEEE}
}

@article{arm2009security,
  title={Security technology building a secure system using trustzone technology (white paper)},
  author={ARM, Architecure},
  journal={ARM Limited},
  year={2009}
}

@article{khalid2022vulnerability,
  title={Vulnerability analysis of Qualcomm Secure Execution Environment (QSEE)},
  author={Khalid, Fatima and Masood, Ammar},
  journal={Computers \& Security},
  pages={102628},
  year={2022},
  publisher={Elsevier}
}

@inproceedings{machiry2017boomerang,
  title={BOOMERANG: Exploiting the Semantic Gap in Trusted Execution Environments.},
  author={Machiry, Aravind and Gustafson, Eric and Spensky, Chad and Salls, Christopher and Stephens, Nick and Wang, Ruoyu and Bianchi, Antonio and Choe, Yung Ryn and Kruegel, Christopher and Vigna, Giovanni},
  booktitle={NDSS},
  year={2017}
}

@misc{Qualcomm2017PA,
  title={Pointer Authentication on ARMv8.3},
  author={Qualcomm},
  year={2017},
  note={\url{https://www.qualcomm.com/content/dam/qcomm-martech/dm-assets/documents/pointer-auth-v7.pdf}}
}

@misc{ARMFVP,
  title={Fixed Virtual Platforms},
  author={ARM},
  year={2023},
  note={\url{https://developer.arm.com/downloads/-/arm-ecosystem-models}}
}

@misc{OPTEExtest,
  title={OP-TEE sanity testsuite},
  author={OP-TEE},
  year={2023},
  note={\url{https://github.com/OP-TEE/optee\_test}}
}

@misc{ARM2022exception,
  title={Learn the architecture - AArch64 Exception Model},
  author={ARM},
  year={2022},
  note={\url{https://developer.arm.com/documentation/102412/latest}}
}

@article{avanzi2017qarma,
  title={The QARMA block cipher family. Almost MDS matrices over rings with zero divisors, nearly symmetric even-mansour constructions with non-involutory central rounds, and search heuristics for low-latency s-boxes},
  author={Avanzi, Roberto},
  journal={IACR Transactions on Symmetric Cryptology},
  pages={4--44},
  year={2017}
}

@inproceedings{suciu2020horizontal,
  title={Horizontal privilege escalation in trusted applications},
  author={Suciu, Darius and McLaughlin, Stephen and Simon, Laurent and Sion, Radu},
  booktitle={Proceedings of the 29th USENIX Conference on Security Symposium},
  pages={825--840},
  year={2020}
}

@misc{GlobalPlatform2021API,
  title={TEE Internal Core API Specification Version 1.3.1},
  author={GlobalPlatform},
  year={2021},
  note={\url{https://globalplatform.org/specs-library/tee-internal-core-api-specification}}
}

@misc{GlobalPlatform2010ClientAPI,
  title={TEE Client API Specification Version 1.0},
  author={GlobalPlatform},
  year={2010},
  note={\url{https://globalplatform.org/wp-content/uploads/2010/07/TEE_Client_API_Specification-V1.0.pdf}}
}

@inproceedings{ravichandran2022pacman,
  title={PACMAN: attacking ARM pointer authentication with speculative execution},
  author={Ravichandran, Joseph and Na, Weon Taek and Lang, Jay and Yan, Mengjia},
  booktitle={Proceedings of the 49th Annual International Symposium on Computer Architecture},
  pages={685--698},
  year={2022}
}

@inproceedings{liljestrand2019pac,
  title={PAC it up: Towards Pointer Integrity using ARM Pointer Authentication.},
  author={Liljestrand, Hans and Nyman, Thomas and Wang, Kui and Perez, Carlos Chinea and Ekberg, Jan-Erik and Asokan, N},
  booktitle={USENIX Security Symposium},
  pages={177--194},
  year={2019}
}

@inproceedings{yoo2022kernel,
  title={In-Kernel Control-Flow Integrity on Commodity OSes using ARM Pointer Authentication},
  author={Yoo, Sungbae and Park, Jinbum and Kim, Seolheui and Kim, Yeji and Kim, Taesoo},
  booktitle={31st USENIX Security Symposium (USENIX Security 22)},
  pages={89--106},
  year={2022}
}

@inproceedings{li2022pacmem,
  title={PACMem: Enforcing Spatial and Temporal Memory Safety via ARM Pointer Authentication},
  author={Li, Yuan and Tan, Wende and Lv, Zhizheng and Yang, Songtao and Payer, Mathias and Liu, Ying and Zhang, Chao},
  booktitle={Proceedings of the 2022 ACM SIGSAC Conference on Computer and Communications Security},
  pages={1901--1915},
  year={2022}
}

@inproceedings{farkhani2021ptauth,
  title={PTAuth: Temporal Memory Safety via Robust Points-to Authentication.},
  author={Farkhani, Reza Mirzazade and Ahmadi, Mansour and Lu, Long},
  booktitle={USENIX Security Symposium},
  pages={1037--1054},
  year={2021}
}

@inproceedings{serra2022pac,
  title={PAC-PL: Enabling control-flow integrity with pointer authentication in FPGA SoC platforms},
  author={Serra, Gabriele and Fara, Pietro and Cicero, Giorgiomaria and Restuccia, Francesco and Biondi, Alessandro},
  booktitle={2022 IEEE 28th Real-Time and Embedded Technology and Applications Symposium (RTAS)},
  pages={241--253},
  year={2022},
  organization={IEEE}
}

@inproceedings{schilling2022fipac,
  title={FIPAC: Thwarting Fault-and Software-Induced Control-Flow Attacks with ARM Pointer Authentication},
  author={Schilling, Robert and Nasahl, Pascal and Mangard, Stefan},
  booktitle={Constructive Side-Channel Analysis and Secure Design: 13th International Workshop, COSADE 2022, Leuven, Belgium, April 11-12, 2022, Proceedings},
  pages={100--124},
  year={2022},
  organization={Springer}
}

@article{yang2019arm,
  title={ARM pointer authentication based forward-edge and backward-edge control flow integrity for kernels},
  author={Yang, Yutian and Zhu, Songbo and Shen, Wenbo and Zhou, Yajin and Sun, Jiadong and Ren, Kui},
  journal={arXiv preprint arXiv:1912.10666},
  year={2019}
}

@inproceedings{denis2020camouflage,
  title={Camouflage: Hardware-assisted cfi for the arm linux kernel},
  author={Denis-Courmont, R{\'e}mi and Liljestrand, Hans and Chinea, Carlos and Ekberg, Jan-Erik},
  booktitle={2020 57th ACM/IEEE Design Automation Conference (DAC)},
  pages={1--6},
  year={2020},
  organization={IEEE}
}

@inproceedings{ismail2022tightly,
  title={Tightly Seal Your Sensitive Pointers with PACTight},
  author={Ismail, Mohannad and Quach, Andrew and Jelesnianski, Christopher and Jang, Yeongjin and Min, Changwoo},
  booktitle={31st USENIX Security Symposium (USENIX Security 22)},
  pages={3717--3734},
  year={2022}
}

@inproceedings{liljestrand2021pacstack,
  title={PACStack: an Authenticated Call Stack.},
  author={Liljestrand, Hans and Nyman, Thomas and Gunn, Lachlan J and Ekberg, Jan-Erik and Asokan, N},
  booktitle={USENIX Security Symposium},
  pages={357--374},
  year={2021}
}

@inproceedings{wang2022rettag,
  title={RetTag: Hardware-assisted return address integrity on RISC-V},
  author={Wang, Yu and Wu, Jinting and Yue, Tai and Ning, Zhenyu and Zhang, Fengwei},
  booktitle={Proceedings of the 15th European Workshop on Systems Security},
  pages={50--56},
  year={2022}
}

@inproceedings{kim2020hardware,
  title={Hardware-based always-on heap memory safety},
  author={Kim, Yonghae and Lee, Jaekyu and Kim, Hyesoon},
  booktitle={2020 53rd Annual IEEE/ACM International Symposium on Microarchitecture (MICRO)},
  pages={1153--1166},
  year={2020},
  organization={IEEE}
}

@inproceedings{mckee2022preventing,
  title={Preventing kernel hacks with HAKC},
  author={McKee, Derrick and Giannaris, Yianni and Perez, Carolina Ortega and Shrobe, Howard and Payer, Mathias and Okhravi, Hamed and Burow, Nathan},
  booktitle={Proceedings 2022 Network and Distributed System Security Symposium. NDSS},
  volume={22},
  pages={1--17},
  year={2022}
}

@inproceedings{kimrisc,
  title={AOS-RISC-V: Towards Always-On Heap Memory Safety},
  author={Kim, Yonghae and Kar, Anurag and Singh, Siddant and Ratnani, Ammar A and Lee, Jaekyu and Kim, Hyesoon},
  booktitle={Sixth Workshop on Computer Architecture Research with RISC-V (CARRV 2022) at the 49th International Symposium on Computer Architecture (ISCA-2022)},
  year={2022}
}

@article{rosenberg2014reflections,
  title={Reflections on trusting trustzone},
  author={Rosenberg, Dan},
  journal={BlackHat USA},
  year={2014}
}

@article{shen2015exploiting,
  title={Exploiting trustzone on android},
  author={Shen, Di},
  journal={Black Hat USA},
  volume={2},
  pages={267--280},
  year={2015}
}

@inproceedings{cerdeira2022rezone,
  title={ReZone: Disarming TrustZone with TEE Privilege Reduction},
  author={Cerdeira, David and Martins, Jos{\'e} and Santos, Nuno and Pinto, Sandro},
  booktitle={31st USENIX Security Symposium (USENIX Security 22)},
  pages={2261--2279},
  year={2022}
}

@misc{Beniamini2016CVE,
  title={TrustZone Kernel Privilege Escalation (CVE-2016-2431)},
  author={Beniamini, Gal},
  year={2016},
  note={\url{http://bits-please.blogspot.com/2016/06/trustzone-kernel-privilege-escalation.html}}
}

@misc{berard2018kinibi,
  title={Kinibi tee: Trusted application exploitation},
  author={Berard, David},
  year={2018},
  note={\url{https://www.synacktiv.com/en/publications/kinibi-tee-trusted-application-exploitation.html}}
}

@misc{Komaromy2018Unbox,
  title={Unbox Your Phone},
  author={Komaromy, Daniel},
  year={2018},
  note={\url{https://medium.com/taszksec/unbox-your-phone-part-i-331bbf44c30c}}
}

@inproceedings{brasser2019sanctuary,
  title={SANCTUARY: ARMing TrustZone with User-space Enclaves.},
  author={Brasser, Ferdinand and Gens, David and Jauernig, Patrick and Sadeghi, Ahmad-Reza and Stapf, Emmanuel},
  booktitle={NDSS},
  year={2019}
}

@inproceedings{sun2015trustice,
  title={Trustice: Hardware-assisted isolated computing environments on mobile devices},
  author={Sun, He and Sun, Kun and Wang, Yuewu and Jing, Jiwu and Wang, Haining},
  booktitle={2015 45th Annual IEEE/IFIP International Conference on Dependable Systems and Networks},
  pages={367--378},
  year={2015},
  organization={IEEE}
}

@inproceedings{proskurin2020xmp,
  title={xmp: Selective memory protection for kernel and user space},
  author={Proskurin, Sergej and Momeu, Marius and Ghavamnia, Seyedhamed and Kemerlis, Vasileios P and Polychronakis, Michalis},
  booktitle={2020 IEEE Symposium on Security and Privacy (SP)},
  pages={563--577},
  year={2020},
  organization={IEEE}
}

@misc{Wiklander2018avb,
  title={Android Verified Boot (AVB) in OP-TEE},
  author={Wiklander, Jens},
  year={2018},
  note={\url{https://github.com/OP-TEE/optee_os/tree/master/ta/avb}}
}

@misc{Garg2020trustedkeys,
  title={Trusted keys in OP-TEE},
  author={Garg, Sumit}, 
  year={2020},
  note={\url{https://github.com/OP-TEE/optee_os/tree/master/ta/trusted_keys}}
}

@inproceedings{mo2020darknetz,
  title={Darknetz: towards model privacy at the edge using trusted execution environments},
  author={Mo, Fan and Shamsabadi, Ali Shahin and Katevas, Kleomenis and Demetriou, Soteris and Leontiadis, Ilias and Cavallaro, Andrea and Haddadi, Hamed},
  booktitle={Proceedings of the 18th International Conference on Mobile Systems, Applications, and Services},
  pages={161--174},
  year={2020}
}

@inproceedings{li2021twinvisor,
  title={Twinvisor: Hardware-isolated confidential virtual machines for arm},
  author={Li, Dingji and Mi, Zeyu and Xia, Yubin and Zang, Binyu and Chen, Haibo and Guan, Haibing},
  booktitle={Proceedings of the ACM SIGOPS 28th Symposium on Operating Systems Principles},
  pages={638--654},
  year={2021}
}

@article{wang2023cacheiee,
  title={CacheIEE: Cache-assisted Isolated Execution Environment on ARM Multi-Core Platforms},
  author={Wang, Jie and Sun, Kun and Lei, Lingguang and Wang, Yuewu and Jing, Jiwu and Wan, Shengye and Li, Qi},
  journal={IEEE Transactions on Dependable and Secure Computing},
  year={2023},
  publisher={IEEE}
}

@inproceedings{yun2019ginseng,
  title={Ginseng: Keeping Secrets in Registers When You Distrust the Operating System.},
  author={Yun, Min Hong and Zhong, Lin},
  booktitle={NDSS},
  year={2019}
}

@misc{2023hafnium,
  title={Hafnium - Trusted Firmware},
  author={Trusted Firmware}, 
  year={2023},
  note={\url{https://www.trustedfirmware.org/projects/hafnium/}}
}

@misc{Apple2010Platform,
  title={Apple Platform Security},
  author={Apple},
  year={2022},
  note={\url{https://help.apple.com/pdf/security/en_US/apple-platform-security-guide.pdf}}
}

@inproceedings{yitbarek2017cold,
  title={Cold boot attacks are still hot: Security analysis of memory scramblers in modern processors},
  author={Yitbarek, Salessawi Ferede and Aga, Misiker Tadesse and Das, Reetuparna and Austin, Todd},
  booktitle={2017 IEEE International Symposium on High Performance Computer Architecture (HPCA)},
  pages={313--324},
  year={2017},
  organization={IEEE}
}

@inproceedings{lee2020off,
  title={An $\{$Off-Chip$\}$ attack on hardware enclaves via the memory bus},
  author={Lee, Dayeol and Jung, Dongha and Fang, Ian T and Tsai, Chia-Che and Popa, Raluca Ada},
  booktitle={29th USENIX Security Symposium (USENIX Security 20)},
  year={2020}
}

@inproceedings{watson2015cheri,
  title={CHERI: A hybrid capability-system architecture for scalable software compartmentalization},
  author={Watson, Robert NM and Woodruff, Jonathan and Neumann, Peter G and Moore, Simon W and Anderson, Jonathan and Chisnall, David and Dave, Nirav and Davis, Brooks and Gudka, Khilan and Laurie, Ben and others},
  booktitle={2015 IEEE Symposium on Security and Privacy},
  pages={20--37},
  year={2015},
  organization={IEEE}
}

@inproceedings{joannou2017efficient,
  title={Efficient tagged memory},
  author={Joannou, Alexandre and Woodruff, Jonathan and Kovacsics, Robert and Moore, Simon W and Bradbury, Alex and Xia, Hongyan and Watson, Robert NM and Chisnall, David and Roe, Michael and Davis, Brooks and others},
  booktitle={2017 IEEE International Conference on Computer Design (ICCD)},
  pages={641--648},
  year={2017},
  organization={IEEE}
}

@article{woodruff2014cheri,
  title={The CHERI capability model: Revisiting RISC in an age of risk},
  author={Woodruff, Jonathan and Watson, Robert NM and Chisnall, David and Moore, Simon W and Anderson, Jonathan and Davis, Brooks and Laurie, Ben and Neumann, Peter G and Norton, Robert and Roe, Michael},
  journal={ACM SIGARCH Computer Architecture News},
  volume={42},
  number={3},
  pages={457--468},
  year={2014},
  publisher={ACM New York, NY, USA}
}

@article{rosenberg2013unlocking,
  title={Unlocking the motorola bootloader},
  author={Rosenberg, Dan},
  journal={Azimuth Security Blog},
  year={2013}
}

@inproceedings{tang2017clkscrew,
  title={$\{$CLKSCREW$\}$: Exposing the perils of $\{$Security-Oblivious$\}$ energy management},
  author={Tang, Adrian and Sethumadhavan, Simha and Stolfo, Salvatore},
  booktitle={26th USENIX Security Symposium (USENIX Security 17)},
  pages={1057--1074},
  year={2017}
}

@inproceedings{qiu2019voltjockey,
  title={VoltJockey: Breaching TrustZone by software-controlled voltage manipulation over multi-core frequencies},
  author={Qiu, Pengfei and Wang, Dongsheng and Lyu, Yongqiang and Qu, Gang},
  booktitle={Proceedings of the 2019 ACM SIGSAC Conference on Computer and Communications Security},
  pages={195--209},
  year={2019}
}

@inproceedings{van2016drammer,
  title={Drammer: Deterministic rowhammer attacks on mobile platforms},
  author={Van Der Veen, Victor and Fratantonio, Yanick and Lindorfer, Martina and Gruss, Daniel and Maurice, Cl{\'e}mentine and Vigna, Giovanni and Bos, Herbert and Razavi, Kaveh and Giuffrida, Cristiano},
  booktitle={Proceedings of the 2016 ACM SIGSAC conference on computer and communications security},
  pages={1675--1689},
  year={2016}
}

@inproceedings{yu2023capstone,
  title={Capstone: A Capability-based Foundation for Trustless Secure Memory Access},
  author={Yu, Jason Zhijingcheng and Watt, Conrad and Badole, Aditya and Carlson, Trevor E and Saxena, Prateek},
  booktitle={32nd USENIX Security Symposium (USENIX Security 23)},
  pages={787--804},
  year={2023}
}

@inproceedings{sehr2010adapting,
  title={Adapting Software Fault Isolation to Contemporary $\{$CPU$\}$ Architectures},
  author={Sehr, David and Muth, Robert and Biffle, Cliff and Khimenko, Victor and Pasko, Egor and Schimpf, Karl and Yee, Bennet and Chen, Brad},
  booktitle={19th USENIX Security Symposium (USENIX Security 10)},
  year={2010}
}

@inproceedings{morrisett2012rocksalt,
  title={RockSalt: better, faster, stronger SFI for the x86},
  author={Morrisett, Greg and Tan, Gang and Tassarotti, Joseph and Tristan, Jean-Baptiste and Gan, Edward},
  booktitle={Proceedings of the 33rd ACM SIGPLAN conference on Programming Language Design and Implementation},
  pages={395--404},
  year={2012}
}

@inproceedings{narayan2020retrofitting,
  title={Retrofitting fine grain isolation in the Firefox renderer},
  author={Narayan, Shravan and Disselkoen, Craig and Garfinkel, Tal and Froyd, Nathan and Rahm, Eric and Lerner, Sorin and Shacham, Hovav and Stefan, Deian},
  booktitle={29th USENIX Security Symposium (USENIX Security 20)},
  pages={699--716},
  year={2020}
}

@inproceedings{johnson2021trust,
  title={Trust, but verify: SFI safety for native-compiled Wasm},
  author={Johnson, Evan and Thien, David and Alhessi, Yousef and Narayan, Shravan and Brown, Fraser and Lerner, Sorin and McMullen, Tyler and Savage, Stefan and Stefan, Deian},
  booktitle={Network and Distributed System Security Symposium (NDSS). Internet Society},
  year={2021}
}

@inproceedings{bosamiya2022provably,
  title={$\{$Provably-Safe$\}$ multilingual software sandboxing using $\{$WebAssembly$\}$},
  author={Bosamiya, Jay and Lim, Wen Shih and Parno, Bryan},
  booktitle={31st USENIX Security Symposium (USENIX Security 22)},
  pages={1975--1992},
  year={2022}
}

@inproceedings{hua2017vtz,
  title={$\{$vTZ$\}$: Virtualizing $\{$ARM$\}$$\{$TrustZone$\}$},
  author={Hua, Zhichao and Gu, Jinyu and Xia, Yubin and Chen, Haibo and Zang, Binyu and Guan, Haibing},
  booktitle={26th USENIX Security Symposium (USENIX Security 17)},
  pages={541--556},
  year={2017}
}

@inproceedings{cho2016hardware,
  title={$\{$Hardware-Assisted$\}$$\{$On-Demand$\}$ Hypervisor Activation for Efficient Security Critical Code Execution on Mobile Devices},
  author={Cho, Yeongpil and Shin, Junbum and Kwon, Donghyun and Ham, MyungJoo and Kim, Yuna and Paek, Yunheung},
  booktitle={2016 USENIX Annual Technical Conference (USENIX ATC 16)},
  pages={565--578},
  year={2016}
}

@article{jang2016privatezone,
  title={Privatezone: Providing a private execution environment using arm trustzone},
  author={Jang, Jinsoo and Choi, Changho and Lee, Jaehyuk and Kwak, Nohyun and Lee, Seongman and Choi, Yeseul and Kang, Brent Byunghoon},
  journal={IEEE Transactions on Dependable and Secure Computing},
  volume={15},
  number={5},
  pages={797--810},
  year={2016},
  publisher={IEEE}
}

@article{carter1994hardware,
  title={Hardware support for fast capability-based addressing},
  author={Carter, Nicholas P and Keckler, Stephen W and Dally, William J},
  journal={ACM SIGOPS Operating Systems Review},
  volume={28},
  number={5},
  pages={319--327},
  year={1994},
  publisher={ACM New York, NY, USA}
}

@inproceedings{filardo2020cornucopia,
  title={Cornucopia: Temporal safety for CHERI heaps},
  author={Filardo, Nathaniel Wesley and Gutstein, Brett F and Woodruff, Jonathan and Ainsworth, Sam and Paul-Trifu, Lucian and Davis, Brooks and Xia, Hongyan and Napierala, Edward Tomasz and Richardson, Alexander and Baldwin, John and others},
  booktitle={2020 IEEE Symposium on Security and Privacy (SP)},
  pages={608--625},
  year={2020},
  organization={IEEE}
}

@incollection{samarati2000access,
  title={Access control: Policies, models, and mechanisms},
  author={Samarati, Pierangela and de Vimercati, Sabrina Capitani},
  booktitle={International school on foundations of security analysis and design},
  pages={137--196},
  year={2000},
  publisher={Springer}
}

@article{hardy1988confused,
  title={The Confused Deputy: (or why capabilities might have been invented)},
  author={Hardy, Norm},
  journal={ACM SIGOPS Operating Systems Review},
  volume={22},
  number={4},
  pages={36--38},
  year={1988},
  publisher={ACM New York, NY, USA}
}

@misc{MITRE2023cwe,
  title={Common weakness enumeration.},
  author={MITRE},
  year={2023},
  note={\url{https://cwe.mitre.org/}}
}

\begin{IEEEbiography} [{\includegraphics[width=1in,height=1.25in,clip,keepaspectratio]{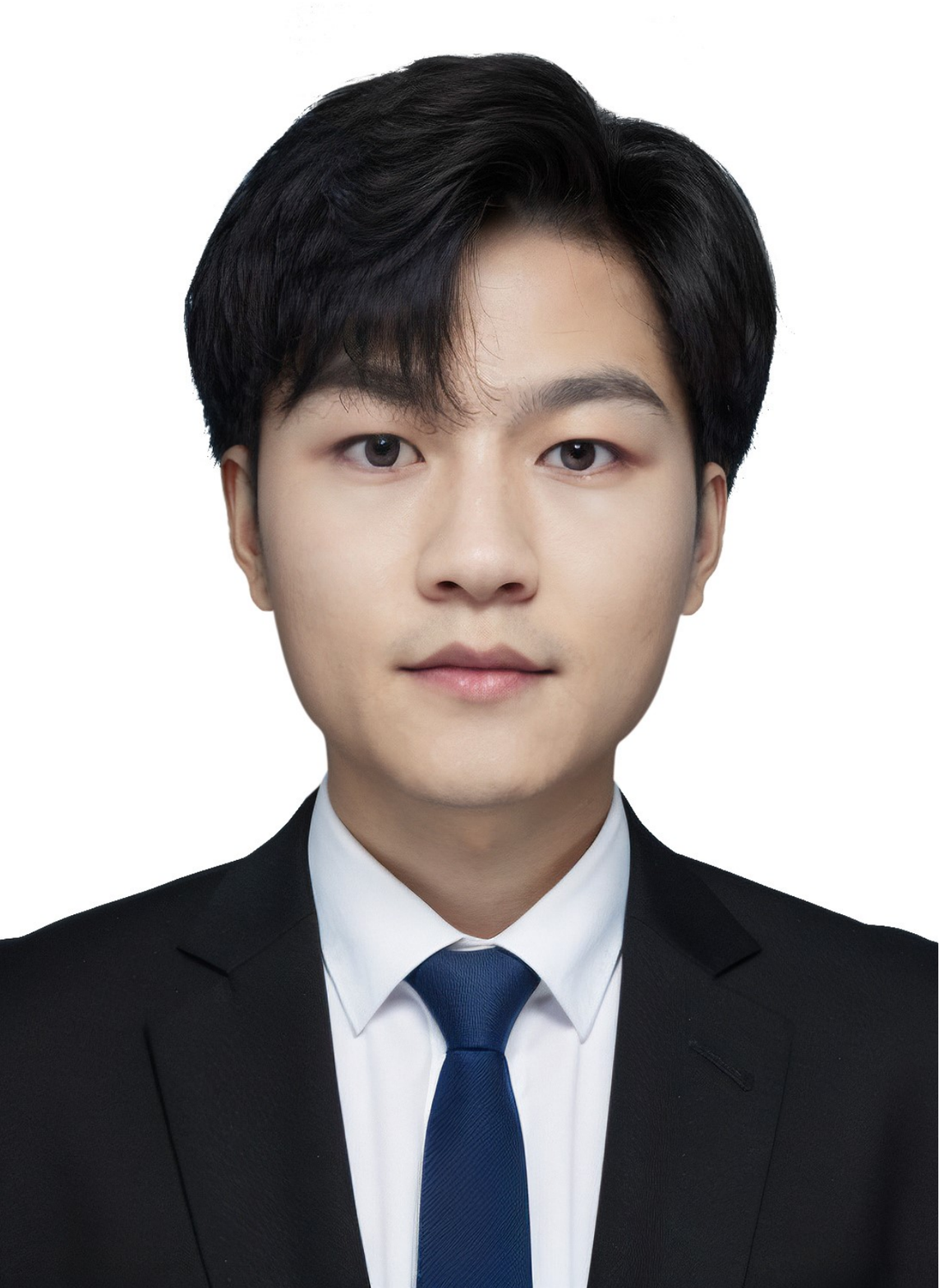}}]
{Shiqi Liu}
received his BS degree in the School of Cyber Science and Engineering from the University of International Relations, Beijing, China, in 2022. He is currently working toward an MS degree with the School of Cyber Science and Engineering, Huazhong University of Science and Technology, Wuhan, China. His research interests include trusted computing and system security.
\end{IEEEbiography}

\begin{IEEEbiography} [{\includegraphics[width=1in,height=1.25in,clip,keepaspectratio]{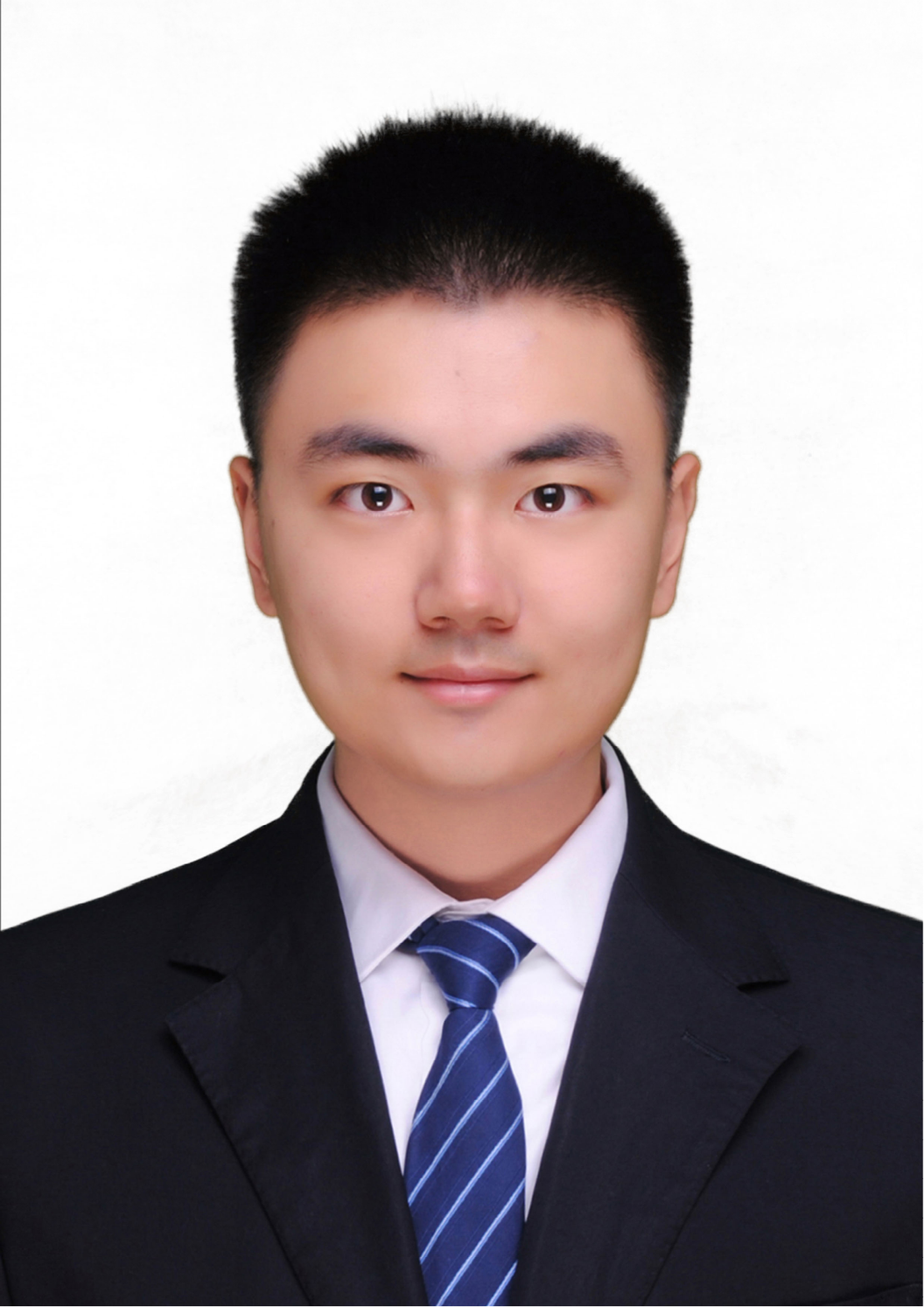}}]
{Xiang Li}
received his MS degree in the School of Cyber Engineering, Xidian University, Xi'an, China, in 2022. He has been a Research Assistant at Southern University of Science and Technology (SUSTech). He is currently a Senior Research Assistant at the Research Center for Basic Theories of Intelligent Computing, Research Institute of Basic Theories, Zhejiang Laboratory, Hangzhou, China. His research interests include confidential computing and system security.
\end{IEEEbiography}

\begin{IEEEbiography} [{\includegraphics[width=1in,height=1.25in,clip,keepaspectratio]{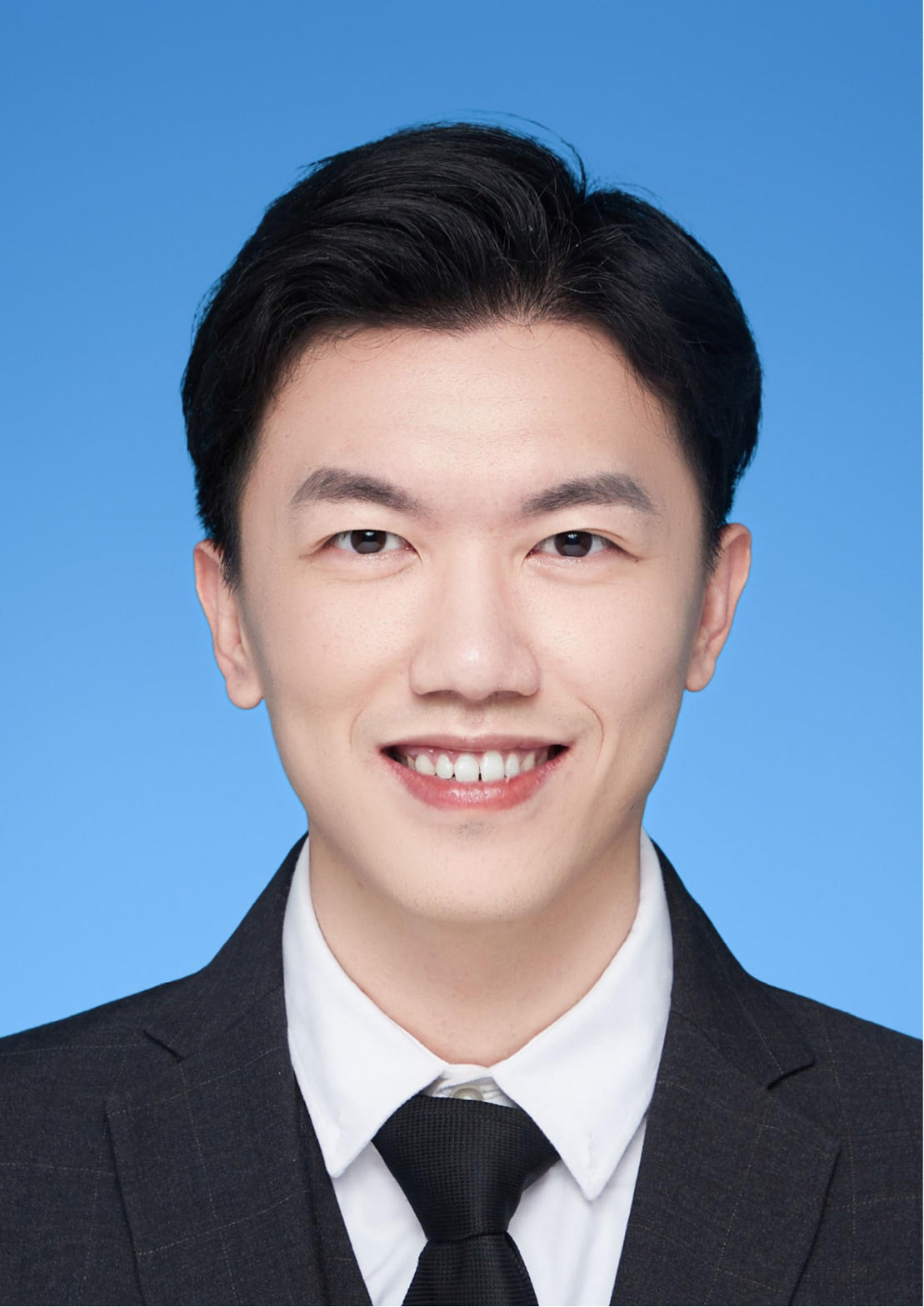}}]
{Jie Wang}
received his Ph.D. degree from the University of Chinese Academy of Sciences (UCAS) in 2021. He has been a Visiting Scholar at George Mason University. He is currently an Associate Professor at Huazhong University of Science and Technology. His research interests are mobile security, trusted computing, and hardware security.
\end{IEEEbiography} 

\begin{IEEEbiography} [{\includegraphics[width=1in,height=1.25in,clip,keepaspectratio]{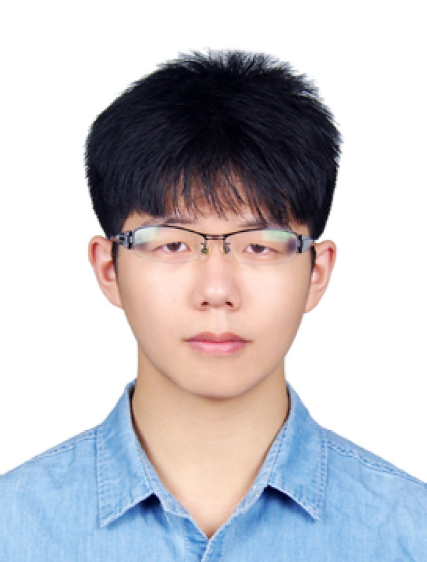}}]
{Yongpeng Gao}
received his BS degree in the School of Cyber Science and Engineering, Huazhong University of Science and Technology, Wuhan, China, in 2023. He is currently working toward an MS degree at the same institution.
\end{IEEEbiography} 

\begin{IEEEbiography} [{\includegraphics[width=1in,height=1.25in,clip,keepaspectratio]{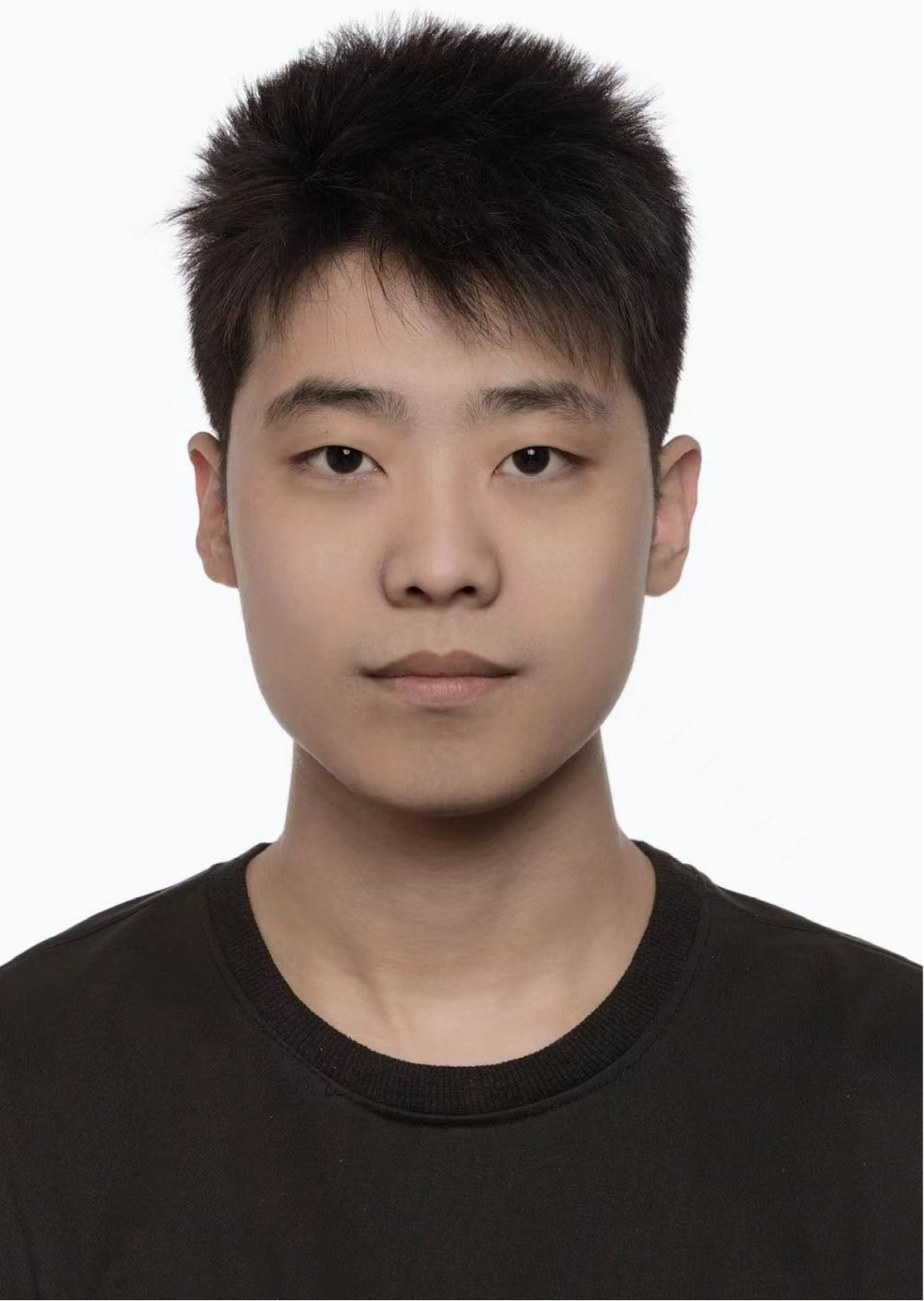}}]
{Jiajin Hu}
received his BS degree in the School of Cyber Science and Engineering, Huazhong University of Science and Technology, Wuhan, China, in 2023. He is currently working toward an MS degree at the same institution.
\end{IEEEbiography}

\end{document}